\documentclass[12pt,preprint]{aastex}

\newcommand\lsim{\mathrel{\rlap{\lower4pt\hbox{\hskip1pt$\sim$}}
        \raise1pt\hbox{$<$}}}
\newcommand\gsim{\mathrel{\rlap{\lower4pt\hbox{\hskip1pt$\sim$}}
        \raise1pt\hbox{$>$}}}

\newcommand{\K}{\mathrm{K}}

    \def\beq{\begin{equation} }
    \def\eeq{\end{equation} }
    \def\spose#1{\hbox to 0pt{#1\hss}}
    \def\ltsim{\mathrel{\spose{\lower.5ex\hbox{$\mathchar"218$}}
     \raise.4ex\hbox{$\mathchar"13C$}}}

\def\spose#1{\hbox to 0pt{#1\hss}}
\def\lta{\mathrel{\spose{\lower 3pt\hbox{$\mathchar"218$}}
        \raise 2.0pt\hbox{$\mathchar"13C$}}}
\def\gta{\mathrel{\spose{\lower 3pt\hbox{$\mathchar"218$}}
        \raise 2.0pt\hbox{$\mathchar"13E$}}}

\shorttitle{Hot Accretion With Outflows}
\shortauthors{Tanaka \& Menou}

\begin{document}

\title{Hot Accretion With Conduction: Spontaneous Thermal Outflows}

\author{Takamitsu Tanaka}
\affil{Department of Astronomy, Columbia University, 550 West
120th Street, New York, NY 10027}
\author{Kristen Menou}
\affil{Department of Astronomy, Columbia University, 550 West
120th Street, New York, NY 10027}

\begin{abstract}
Motivated by the low-collisionality of gas accreted onto black holes
in Sgr~A* and other nearby galactic nuclei, we study a family of 2D
advective accretion solutions with thermal conduction. While we only
impose global inflow, the accretion flow spontaneously develops
bipolar outflows. The role of conduction is key in providing the extra
degree of freedom (latitudinal energy transport) necessary to launch
these rotating thermal outflows.  The sign of the Bernoulli constant
does not discriminate between inflowing and outflowing regions. Our
parameter survey covers mass outflow rates from $\sim 0$ to $13\%$ of
the net inflow rate, outflow velocities from $\sim 0$ to $11\%$ of the
local Keplerian velocity and outflow opening angles from $\sim 0$ to $
60$~degs.  As the magnitude of conduction is increased, outflows can
adopt a conical geometry, pure inflow solutions emerge, and the limit
of 2D non-rotating Bondi-like solutions is eventually reached.  These
results confirm that radiatively-inefficient, hot accretion flows have
a hydrodynamical propensity to generate bipolar thermal outflows.
\end{abstract}


\section{Introduction}

Over the past decade, X-ray observations have made it increasingly
clear that black holes are capable of accreting gas under a variety of
flow configurations. In particular, there is convincing evidence that
a distinct form of hot accretion exists at sub-Eddington rates, which
contrasts with the classical ``cold and thin'' accretion disk scenario
(Shakura \& Sunyaev 1973). Hot accretion appears to be common in the
population of supermassive black holes in galactic nuclei and during
the quiescent phases of accretion onto stellar-mass black holes in
X-ray transients (e.g. Narayan et al. 1998a; Lasota et al. 1996; Di
Matteo et al. 2000; Esin et al. 1997, 2001; Menou et al. 1999; see
Narayan, Mahadevan \& Quataert, 1998b; Melia \& Falcke 2001; Narayan
2003; Narayan \& Quataert 2005 for reviews).

Theories for the structure and properties of hot accretion flows
remain controversial, however.  Inspired by the work of Shapiro,
Lightman \& Eardley (1976), Narayan \& Yi (1994; 1995a,b) derived
self-similar ADAF solutions which emphasized the important stabilizing
role of radial heat advection (see also Ichimaru 1977; Rees et
al. 1982; Abramowicz et al. 1988).  Subsequent analytical work on hot
accretion flows has emphasized the possibility of outflows, motivated
by a positive Bernoulli constant (Narayan \& Yi 1994; ADIOS: Blandford
\& Begelman 1999), and the potential role of convection (CDAFs;
Narayan et al. 2000, Quataert \& Gruzinov 2000a). Hydrodynamical and
MHD numerical simulations, on the other hand, have greatly contributed
to the subject by highlighting important dynamical aspects of the
problem which are not captured by idealized analytical models
(e.g. Stone et al. 1999; Stone \& Pringle 2001; Hawley et al. 2001;
DeVilliers et al. 2003; Proga \& Begelman 2003; McKinney \& Gammie
2004).

The purpose of this work is to consider the possibility that thermal
conduction, which has been a largely neglected ingredient, could
affect the global properties of hot accretion flows substantially. In
\S2, we use existing observational constraints on a few nearby
galactic nuclei to argue that hot accretion is likely to proceed under
weakly-collisional conditions in these systems, thus implying that
thermal conduction could be important. In \S3, we extend the original
1D self-similar solutions of Narayan \& Yi (1994) to include a
saturated form of thermal conduction and we study the effects of
conduction on the flow structure and properties. In \S4, we generalize
the 2D solutions of Narayan \& Yi (1995a) in the presence of thermal
conduction and find that this extra degree of freedom allows the
emergence of rotating thermal outflows. In \S5, we discuss our results
in the context of recently established observational trends for hot
accretion flows and related outflows.  In \S6, we comment on several
limitations of our work, potential additional consequences of strong
thermal conduction in hot accretion flows as well as directions for
future work.

\section{Observational Constraints and Collisionality}

{\it Chandra} observations provide tight constraints on the density
and temperature of gas at or near the Bondi capture radius in Sgr A*
and several other nearby galactic nuclei. These observational
constraints have been used before to estimate the rate at which the
gas is captured by the black hole in these systems, following Bondi
theory (e.g. Loewenstein et al. 2001; Di Matteo et al. 2003; Narayan
2003). Here, we use these same constraints (taken from Loewenstein et
al. 2001; Baganoff et al. 2003; Di Matteo et al. 2003; Ho, Terashima
\& Ulvestad 2003) to calculate mean free paths for the observed gas. It
should be noted that these constraints are not all equally good, in
the sense that {\it Chandra} has probably resolved the gas capture
radius in Sgr A* and M87 but not in the other nuclei.

Galactic nuclei and their associated gas properties on $1''$ scales
are listed in Table~\ref{tab:one}: $n_{1''}$ is the gas number
density, $T_{1''}$ is the temperature, $R_1$ is the $1''$
size-equivalent at the nucleus distance, $l_1$ is the mean free path
for the observed $1''$ conditions and $R_{\rm cap}$ is the capture
radius, inferred from the gas temperature and the black hole mass in
each nucleus. From Spitzer's (1962) expressions for collision times, a
standard expression for the thermal speed and a Coulomb logarithm $\ln
\Lambda \simeq 20$, one gets the simple mean free path scaling $l
\simeq 10^4 (T^2/n)$~cm (Cowie \& McKee 1997; valid for both electrons
and protons in a 1-temperature gas). We have used this relation to
calculate $l_1$ values in Table~\ref{tab:one}.

The inferred mean free paths are in the few hundredths to few tenths
of the observed $1''$ scales ($l_1/R_1$ values in
Table~\ref{tab:one}). Assuming no change in the gas properties down to
the inferred capture radius, values of $l_1/R_{\rm cap} \gsim 0.1$ are
deduced in four of the six nuclei. One could also reformulate these
scalings by stating that the $1''$ mean free paths are systematically
$\gsim 10^{4-5} R_{\rm s}$, where $R_{\rm s}$ is the Schwarzschild
radius (the typical length-scale associated with the accreting black
holes). These numbers suggest that accretion will proceed
under weakly-collisional conditions in these systems.

The weakly-collisional nature of ADAFs has been noted before
(e.g. Mahadevan \& Quataert 1997), in the sense of collision times
longer than the gas inflow time, but these are model-dependent
statements. Direct observational constraints on the gas properties
near the capture radius favor a weakly-collisional regime of hot
accretion, more or less independently of the exact hot flow
structure. Whether the gas adopts an ADAF, CDAF or ADIOS type
configuration once it crosses the capture radius, it is expected to
become even more weakly-collisional as it approaches the black hole,
since the relative mean free path scales as $l/R \propto R^{-3/2-p}$
in these flow solutions with a virial temperature profile and a
density profile $\rho \propto R^{-3/2+p}$.

Tangled magnetic fields in the turbulent accretion flow would likely
reduce the effective mean free paths of particles. The magnitude of
this reduction, which will depend on the field geometry, is currently
unknown. We return to this important issue in our discussion
section. Even if the effective mean free paths of electrons are
reduced to values $l_{\rm eff}/R \lta 1$, thermal conduction may be
important . It is therefore of interest to investigate its effects on
the structure and properties of hot accretion flows in general. We
begin with a study of 1D ADAF models with conduction in \S3 and then
move on to study a 2D version of these hot accretion flows in \S4.

\section{One-Dimensional Hot Accretion Flows with Saturated Conduction}

\subsection{The Self-Similar Equations in One Dimension}

The standard Spitzer formula for thermal conduction applies only to
gas well into the collisional (``fluid'') regime, with a mean free
path $l \ll L$, for any relevant flow scale $L$. We expect radial
temperature variations on scales $\sim R$ from self-similarity. In a
weakly-collisional regime with $l \gg R$ or even $l \sim R$, a
saturated (or equiv.  ``flux-limited'') thermal conduction formalism
should be adopted if one is to avoid unphysically large heat
fluxes. We adopt the formulation of Cowie and McKee (1977) and write
the saturated conduction flux as $F_s = 5 \Phi_{\rm s} \rho c_s^3$, where
$\Phi_{\rm s}$ is the saturation constant (presumably $\sim 1$), $\rho$ is
the gas mass density and $c_s$ is its isothermal sound speed. With
this prescription, the largest achievable conduction flux is
approximated as the product of the thermal energy density in electrons
times their characteristic thermal velocity (assuming a thermal
distribution and equal electron and ion temperatures; see Cowie and
McKee 1977 for details). Because it is a saturated flux, it no longer
explicitly depends on the magnitude of the temperature gradient but
only on the direction of this gradient. Heat will flow outward in a 1D
hot accretion flow with a near-virial temperature profile, hence the
positive sign adopted for $F_s$.

With this formulation for conduction, the steady-state 1-temperature
self-similar ADAF solution of Narayan \& Yi (1994) can be generalized
to include the divergence of the saturated conduction flux in the
entropy-energy equation (see also Honma 1996 and Manmoto et al. 2000
for investigations of the role of ``turbulent'' heat transport in
one-dimensional ADAF-like flows).  Adopting the same cylindrical
geometry and notation as Narayan \& Yi (1994), we consider a
``visco-turbulent'', differentially rotating hot flow which satisfies
the following height-integrated equations for the conservation of
momentum and energy
\begin{eqnarray}
v_r \frac{dv_r}{dR} &=& R (\Omega^2-\Omega_K^2)-\frac{1}{\rho}\frac{d}{dR}(\rho c_s^2),\\
v_r \frac{d(\Omega R^2)}{dR} &=& \frac{1}{\rho R H} \frac{d}{dR} \left( \frac{\alpha \rho c_s^2 R^3 H}{\Omega_K} \frac{d \Omega}{dR}\right),\\
2 H \rho v_r T \frac{ds}{dR} & = & f \frac{2 \alpha \rho c_s^2 R^2 H}{\Omega_K} \left( \frac{d \Omega}{dR}\right)^2 - \frac{2H}{R} \frac{d}{dR} (R F_s),\label{eq:ent}
\end{eqnarray}
supplemented by the continuity equation, which is rewritten as $\dot
M= -4 \pi R H v_r \rho$. The last term in equation~(\ref{eq:ent}) is
the additional divergence of the saturated conduction flux.  In these
equations, $R$ is the cylindrical radius, $\Omega_K$ is the Keplerian
angular velocity around the central accreting mass, $v_r$ is the gas
radial inflow speed, $\Omega$ is its angular velocity, $\rho$ is the
mass density, $c_s$ is the isothermal sound speed, $s$ is the gas
specific entropy, $T$ is its temperature, $H= R c_s/v_K$ is the flow
vertical thickness and $f \leq 1$ is the advection parameter ($f=1$
for negligible gas cooling). A Shakura-Sunyaev prescription has been
used to capture the ``visco-turbulent'' nature of the flow, with an
equivalent kinematic viscosity coefficient $\nu =\alpha
c_s^2/\Omega_K$, where $\alpha$ is the traditional viscosity
parameter.

We look for solutions to these equations of the form
\begin{equation}
\rho = \rho_0 R^{-3/2},~v_r=v_0 R^{-1/2},~\Omega=\Omega_0
R^{-3/2},~c_s^2=c_{s0}^2 R^{-1},
\end{equation}
and use the notation $v_K=v_{K0} R^{-1/2}=R \Omega_K=R \Omega_{K0}
R^{-3/2}$.  We find that the three unknowns $v_0$, $\Omega_0$ and
$c_{s0}$ satisfy the relations
\begin{eqnarray}
\frac{1}{2}v_0^2 &-& \Omega_{K0}^2 + \Omega_0^2 + \frac{5}{2}c_{s0}^2=0,\label{eq:bern}\\
v_0 &=& -\frac{3 \alpha}{2} \frac{c_{s0}^2}{v_{K0}},\\
\frac{9 \alpha^2}{8} c_{s0}^4 &+& \left( \frac{5}{2} + \frac{5/3-\gamma}{f(\gamma-1)} \right)c_{s0}^2 v_{K0}^2 -\frac{40 \Phi_{\rm s}}{9 \alpha f}c_{s0} v_{K0}^3 \nonumber \\
& & - v_{K0}^4 = 0,\label{eq:enrel}
\end{eqnarray}
while the density scale, $\rho_0$, is fixed by the accretion rate,
$\dot M$.

Contrary to the original ADAF solution, the energy relation
(Eq.~[\ref{eq:enrel}]) is not a quadratic in $(c_{s0}/v_{K0})^2$
because of the extra saturated conduction term. We solve this fourth
order polynomial with a standard numerical technique (Press et
al. 1992) and discard the two imaginary roots as well as the real root
with negative $c_{s0}$, which is unphysical because of a conduction
flux going up the temperature gradient.

\subsection{Results: Hot Accretion in One Dimension}

\begin{figure*}
\plotone{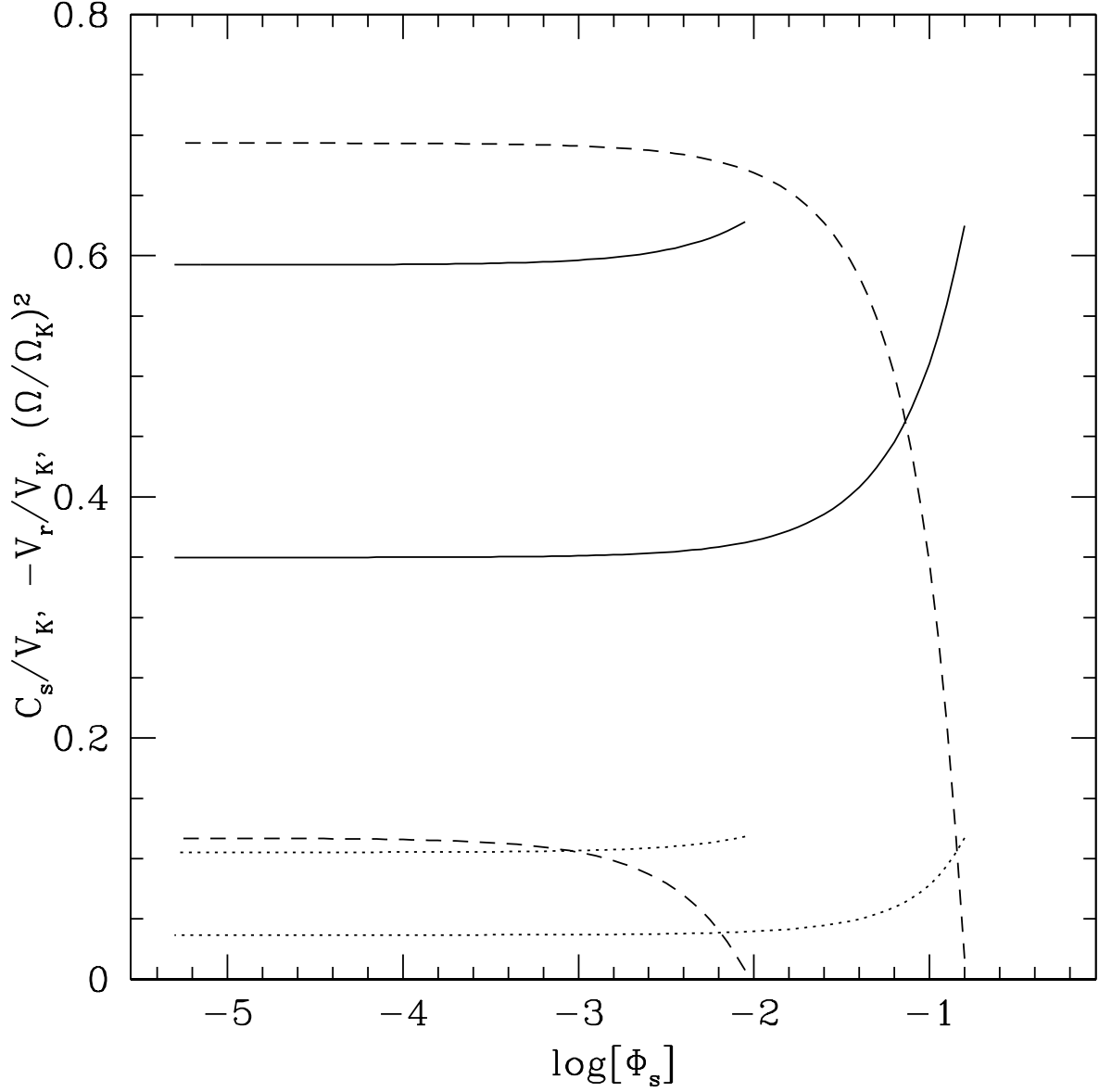}
\caption{\label{fig:one} Variations in the sound speed ($c_s/v_K$,
solid lines), angular velocity squared ($\Omega^2 / \Omega_K^2$,
dashed lines) and radial inflow speed ($- v_r/v_K$, dotted lines) for
two distinct 1D solutions, as a function of the saturation constant,
$\Phi_{\rm s}$, for conduction. At low $\Phi_{\rm s}$ values, these
solutions match the original 1D ADAF. The solution extending up to
$\log \Phi_{\rm s} \simeq -0.8$ has a gas adiabatic index $\gamma =
1.1$, while the solution extending up to $\log \Phi_{\rm s} \simeq -2$
has $\gamma =1.5$.  In both solutions, the viscosity parameter
$\alpha=0.2$ and the advection parameter $f=1$.}
\end{figure*}

The main features of the new self-similar solutions with saturated
conduction are illustrated in Figure~\ref{fig:one}, as a function of
the value of the saturation constant, $\Phi_s$. We are showing results
for two distinct solutions here, with $\gamma=1.5$ and $\gamma=1.1$,
while we have fixed $f=1$ and $\alpha =0.2$ in both cases. Even though
a value of $\gamma < 4/3$ may not be physically meaningful, it is
useful to illustrate more clearly how solutions depend on $\gamma$.

Original 1D ADAF solutions are recovered at small $\Phi_{\rm s}$ values. As
$\Phi_{\rm s}$ approaches unity, however, the solutions deviate
substantially from the standard ADAF. The sound speed, $c_s$, and the
radial inflow speed, $-v_r$, both increase with the magnitude of
conduction, while the squared angular velocity, $\Omega^2$,
decreases. Formal solutions still exist when $\Omega^2$ becomes
negative but they are obviously physically unacceptable. The solution
with $\gamma=1.5$ reaches the non-rotating limit at $\log \Phi_{\rm s}
\simeq -2$, while the solution with $\gamma =1.1$ reaches this limit
for $\log \Phi_{\rm s} \simeq -0.8$. Both solutions have sub-virial
temperatures ($c_s< v_k$) and subsonic radial inflow speeds ($-v_r<
c_s$), which appear to be rather general properties. Supersonic
solutions may exist for large $\alpha$ values when $\gamma \to 5/3$.

As the level of saturated conduction is increased, more and more heat
flows from the hotter, inner regions, resulting in a local increase of
the gas temperature relative to the original ADAF
solution. Simultaneously, the gas adjusts its angular velocity (which
reduces the level of viscous dissipation) and increases its inflow
speed to conserve its momentum balance (increasing at the same time
the level of advection). One can show from the energy equation that
$\Omega^2 \propto (1-|q_{\rm cond}/q_{\rm adv}|)$, where $q_{\rm
cond}$ and $q_{\rm adv}$ are the magnitudes of the conduction and
advection terms: solutions cease to exist once advection balances
exactly the heating due to saturated conduction.  We have found that
variations in the advection parameter, $f$, have relatively minor
effects on the solutions, as long as $f$ is not $\ll 1$. The breakdown
of the solutions (when $\Omega^2 \to 0$) occurs at lower values of the
saturation constant $\Phi_{\rm s}$ for smaller $\alpha$ and larger $\gamma$
values.  Solutions with $\Omega^2 = 0$ can be viewed as Bondi-like
solutions with saturated conduction.

Clearly, conduction can significantly affect the structure and
properties of hot accretion flows, even at a level well below the
saturated flux value (i.e. for $\Phi_{\rm s} \ll 1$).  By analogy with the
reasoning of Narayan \& Yi (1995a) on convection in 2D self-similar
ADAF solutions, one might expect conduction to be more important in
the polar regions of hot accretion flows than at their equator. To
better understand the role of conduction, it is therefore important to
study it in a more realistic two-dimensional setting, without height
integration.

\section{Two-Dimensional Hot Accretion Flows with Thermal Conduction}

\subsection{The Self-Similar Equations in Two Dimensions}
  
We now extend the analysis of the previous section to two-dimensional
hot accretion flows, following closely the methods of Narayan \& Yi
(1995a). We also discuss in Appendix C the relation between our work
and that of Anderson (1987).  Adopting spherical polar coordinates ($r
\theta \phi$), we consider axisymmetric, steady-state flows
($\partial/\partial\phi = \partial/\partial t = 0$). We define
Keplerian angular and linear velocities as, respectively,
\begin{eqnarray}
\Omega_{\K} & \equiv & (GM)^{1/2}r^{-3/2},\\
v_{\K} & \equiv&  \Omega_{\K}r,
\end{eqnarray}
and we use the same $\alpha$-prescription as before to describe the
visco-turbulent nature of the flow:
\begin{equation}
\nu =\alpha c_{s}^{2}/\Omega_{\K},
\end{equation}
where $c_{s}$ is the isothermal sound speed ($c_{s}^{2} \equiv P/\rho$).

The fluid equations obeyed by the visco-turbulent accretion flow are
written as follows:
\begin{eqnarray}
\nabla \cdot (\rho {\bf v}) & \:=\: &0,
\label{cont}\\
({\bf v}\cdot {\bf\nabla}){\bf v} &\:=\:& -\frac{1}{\rho}{\bf\nabla}P - \Omega_{\K}^{2} {\bf r} +
\frac{\alpha c_{s}^{2}}{\Omega_{\K}}  \nabla \cdot {\bf \sigma},\\
{\bf v} \cdot(T{\bf\nabla}s) &\:=\:&f\alpha \frac{ c_{s}^{2}}{\Omega_{\K}}
\left[\frac{1}{2}({\bf \sigma}\cdot{\bf \sigma})-\frac{2}{3}
 (\nabla\cdot{\bf v})^{2}\right]+\frac{1}{\rho}\nabla \cdot (\lambda \nabla T),
\end{eqnarray}
where ${\bf \sigma}$ is the viscous stress tensor (divided by the
dynamic viscosity, $\rho\nu$).  The notation for the other flow
variables is the same as in \S3. It should become clear, later on,
that our main results are not expected to depend critically on the
exact form of the viscous stress tensor adopted. To allow a direct
comparison with our results in \S3 and for consistency with Narayan \&
Yi (1995a), we use the same advection parameter $f$ to quantify the
cooling efficiency of the flow. We will see, however, that this
parameter is in fact ill-defined for solutions with conduction.  Fully
developed forms of these equations (with a few typos and no thermal
conduction term) can be found in Narayan \& Yi (1995a).

Contrary to the formulation adopted in \S3, we do not write the
conduction flux in a saturated form here, but return instead to a
traditional form where the flux depends linearly on the local
temperature gradient. Justifications for this choice are provided in
Appendix~A.  We adopt a form of the thermal conductivity coefficient
which preserves the radial self-similarity of the solutions:
\begin{equation}
\lambda(r) =\lambda_{0}r^{-1}. 
\end{equation}

Following Narayan \& Yi (1995a), we consider separable solutions satisfying
\begin{eqnarray}
v_{r}(r,\theta) &=\: & v_{\K}(r)\,v_r(\theta),\\
v_{\theta}(r,\theta) &=\: &0,\\
v_{\phi}(r,\theta) &=\: & v_{\K}(r)\,\Omega(\theta)\sin\theta , \label{sin}\\
\rho(r,\theta) &= \: & \rho(\theta) \, r^{-3/2},\\
c_{s}(r,\theta) &=\: &v_{\K}(r)\,c_{s}(\theta).
\end{eqnarray}
In Narayan \& Yi (1995a), the $\sin\theta$ in the equivalent of
Eq.~(\ref{sin}) is missing, in what appears to be a typo. Note that we
make the same $v_{\theta} =0$ assumption as Narayan \& Yi (1995a).

With these forms, solutions automatically satisfy the continuity
equation, Eq.~(\ref{cont}). The problem is reduced to a
one-dimensional differential system along the polar coordinate,
$\theta$.  We are left to solve the following four coupled
differential equations for the dimensionless variables $v_r(\theta),
\Omega(\theta), \rho(\theta),$ and $c_{s}(\theta)$:

\begin{eqnarray}
\rho\left(-\frac{1}{2}v_r^{2}-\sin^{2}\theta\Omega^{2}\right) &\:=\:& -\rho +
\frac{5P}{2}+ \alpha P \left(-v_r + \cot \theta \frac{dv_r}{d\theta}\right)
+ \alpha\frac{d}{d\theta}\left(P\frac{dv_r}{d\theta}\right),
\label{moma}\\ \nonumber\\
-\sin\theta\cos\theta\rho\Omega^{2} &=&-\frac{dP}{d\theta} 
+ \alpha \left(v_r\frac{dP}{d\theta}+\frac{3}{2}P\frac{dv_r}{d\theta}\right),
\label{momb}\\\nonumber\\
\frac{1}{2}\sin\theta \rho v_r\Omega &\:=\:&\alpha \left[-\frac{3}{4}P\Omega
+3\cos\theta P\frac{d\Omega}{d\theta}
+\sin\theta\frac{d}{d\theta}\left(P\frac{d\Omega}{d\theta}\right)\right],
\label{momc}\\\nonumber\\
-\frac{3}{2}\epsilon ^{\prime} v_r &\:=\:&\alpha\left\{3v_r^{2}+
\left(\frac{dv_r}{d\theta}\right)^{2}
+\sin^{2}\theta\left[\frac{9}{4}\Omega^{2}
+\left(\frac{d\Omega}{d\theta}\right)^{2}\right] \right\}\nonumber\\
& & \qquad\qquad+\lambda_{0}^{\prime}\frac{1}{P}
\left[c_{s}^{2}+\frac{1}{\sin\theta}\frac{d}{d\theta}
\left(\sin\theta \frac{d c_{s}^{2}}{d\theta}\right)\right].
\label{energy}
\end{eqnarray}

For convenience, the parameters
\begin{eqnarray}
\epsilon ^{\prime} &\equiv&\frac{1}{f}\epsilon=
\frac{1}{f} \frac{5/3 - \gamma}{\gamma - 1},\\
\lambda_{0}^{\prime} &\equiv& \frac{1}{f}\lambda_{0},
\end{eqnarray}
were introduced in the energy equation (Eq.~[\ref{energy}]), where
$\gamma \equiv c_{p}/c_{v}$ is the adiabatic index of the gas.

This is a seventh-order differential system of four variables.  The
three momentum equations (Eqs~[\ref{moma}--\ref{momc}]) are identical
to those presented in Narayan \& Yi (1995a).  The energy equation
(Eq.~[\ref{energy}]) differs in that extra terms corresponding to the
radial and polar contributions to heat conduction are now present.
This qualitative difference turns out to be crucial in providing the
extra degree of energetic freedom necessary for inflow or outflow at
the pole.  While the original energy equation seemed to require that
rotating solutions be static ($v_r=0$) along the polar axis because of
the positivity of all the viscous terms (no rotating solutions with
$v_r\neq 0$ at $\theta=0$ were found by Narayan \& Yi 1995a), the
addition of thermal conduction terms allows for radial velocities at
the pole which are generally non-zero. Furthermore, for outflows to be
energetically favored, the polar contribution to heat conduction
(second term in the last square bracket in Eq.~[\ref{energy}]) must be
negative enough to overcome the positive sum of the radial conduction
term and all the viscous terms on the RHS of Eq.~(\ref{energy}). When
that happens, advection can act as an effective heat source, with a
positive sign for the radial velocity, $v_r$ (on the LHS of
Eq.~[\ref{energy}]). As we shall see below, this sign change for $v_r$
preferentially happens in polar regions of the flow and only for
certain combinations of model parameters.

Boundary conditions for the system of differential equations are
obtained by requiring that all flow variables be symmetric and
continuously differentiable across the pole and the equator. This
corresponds to the following conditions at $\theta = 0$ and $\theta =
\pi/2$:
\begin{equation}
\frac{dv}{d\theta}
=\frac{d\Omega}{d\theta}=\frac{d\rho}{d\theta}=\frac{dc_{s}^2}{d\theta}=0.
\end{equation}
As noticed by Narayan \& Yi (1995a), these conditions are not all
independent from each other, given Eqs.~(\ref{moma}--\ref{energy}), so
that the system ends up not being over-constrained.

The net mass accretion rate, which one obtains from the volume
integral of the continuity equation (Eq.~[\ref{cont}]), is
\begin{equation} \label{eq:mdot}
\dot m =- \int 2 \pi\rho(\theta) v(\theta) \sin \theta d\theta,
\end{equation}
where the radial dependence has been omitted because of
self-similarity.  This integral supplies two additional boundary
conditions to our system because we enforce it by solving the
corresponding differential equation for $d {\dot m}/d \theta$ and
imposing the values of $\dot m$ at $\theta=0$ and $\theta=\pi/2$. The
$\dot m$ integral then uniquely determines the scaling of
$\rho(\theta)$ in our solutions.  In what follows, we have used $\dot
m = 1$ (from $\theta =0$ to $\pi/2$) consistently in all the
solutions.

We find numerical solutions with an implementation of the relaxation
code described in Hameury et al. (1998). We do not implement the
adaptive mesh feature but simply solve the set of equations on a grid
of $400$--$1000$ points equally spaced in $\theta$. After recovering
the original 2D ADAF solutions of Narayan \& Yi (1995a), we explored
the parameter space of solutions with conduction in detail, in terms
of the model parameters $\alpha$, $\epsilon^{\prime}$ and
$\lambda_{0}^{\prime}$.

\subsection{Results: Hot Accretion in Two Dimensions}

\subsubsection{Dynamical Structure of the Flow}

\begin{figure}
\plotone{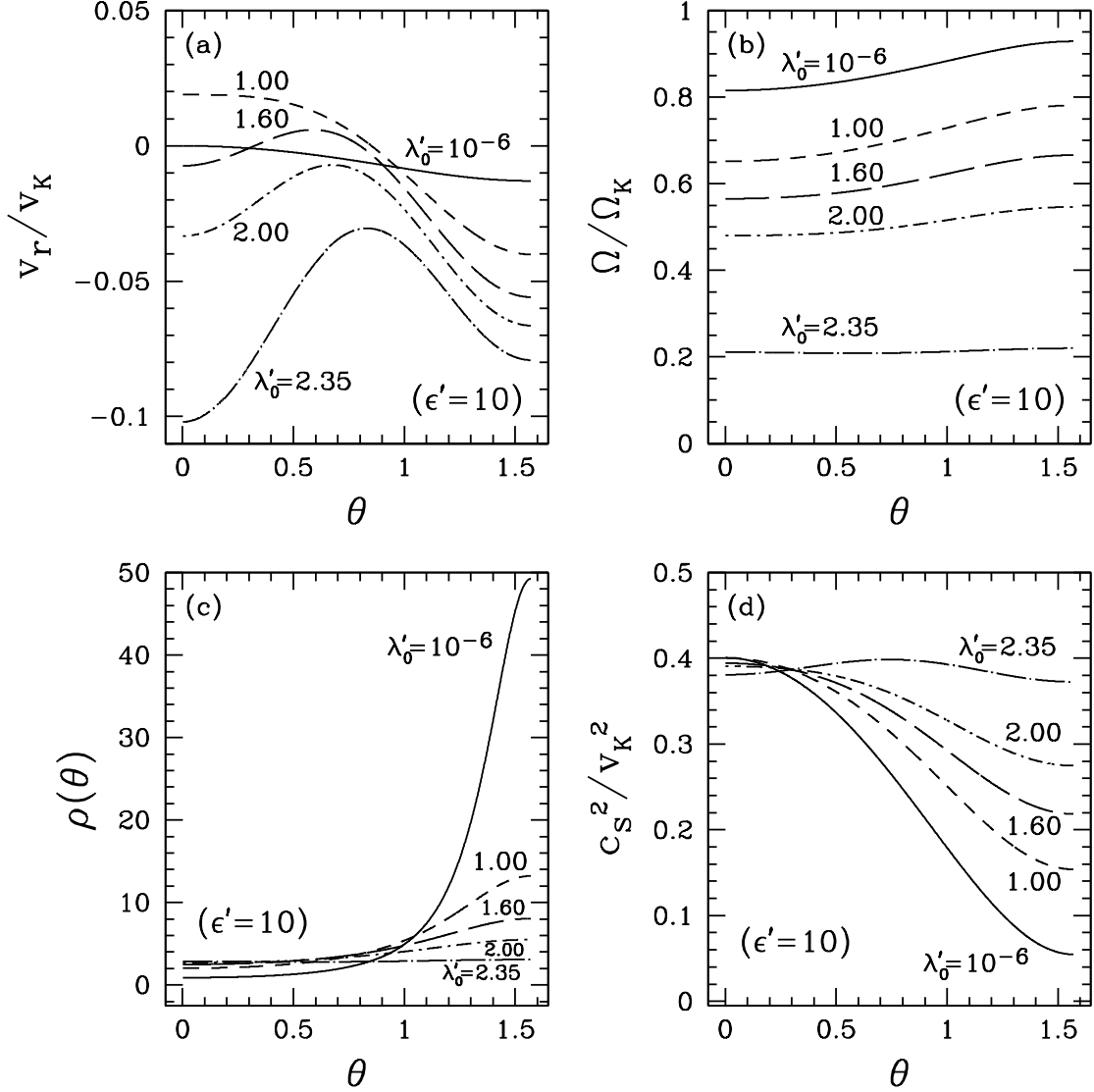}
\caption{Latitudinal profiles of flow dynamical quantities in five 2D
self-similar solutions with $\alpha=0.1$, $\epsilon^{\prime}=10$ and
varying degrees of conduction (as measured by $\lambda_{0}^{\prime}$).
(a): Radial velocity in units of the Keplerian value, $v_r/v_{K}$,
where a positive value indicates outflow. (b): Angular velocity in
units of the Keplerian value, $\Omega/\Omega_{\K}$.  (c): Density,
$\rho(\theta)$, in arbitrary units.  (d): Square of the isothermal
sound speed in units of Keplerian velocity,
$c_{s}{}^{2}/v_{\K}{}^{2}$.}
\label{fig1}
\end{figure}

Figure~\ref{fig1} shows latitudinal profiles ($\theta=0$ at the pole)
of the four dynamical variables in our model, for solutions with
$\alpha=0.1$, $\epsilon^{\prime}=10$ and several values of
$\lambda_{0}^{\prime}$. Except for the density, $\rho$ (which is
normalized to $\dot m =1$), the radial velocity, $v_r$, the angular
velocity, $\Omega$, and the isothermal sound speed squared, $c_s^2$,
are all scaled to the local Keplerian values. For small enough values
of $\lambda_{0}^{\prime}$ (=$10^{-6}$), solutions remain very close to
the original 2D ADAF solutions of Narayan \& Yi (1995a), with $v_r$
(nearly) zero at the pole. Note that this solution, with
$\epsilon^{\prime}=10$, is denser and cooler at the equator than at
the pole and it corresponds to something like a ``thin disk''
surrounded by a hot coronal atmosphere.

As the value of $\lambda_{0}^{\prime}$ is increased, the solutions
start deviating substantially from the original ADAF, with faster
radial inflow at the equator and significant radial outflow appearing
in polar regions. The solution with $\lambda_{0}^{\prime} =1$ in
Fig.~\ref{fig1} illustrates this transition clearly. The flow also
becomes hotter (except around the poles), is rotating more slowly
(being more pressure supported) and has a more uniform density
profile, with a reduced density at the equator: it is gradually
approaching spherical symmetry. As we shall see below, some of these
variations in properties are shared by the original ADAF solutions if
the flow is forced to be more advective (by reducing the value of
$\epsilon^{\prime}$).

As the value of $\lambda_{0}^{\prime}$ is increased further, another
qualitative change in the solutions occurs, with inflow at the pole,
outflowing regions away from the pole and still faster inflow at the
equator. In other words, the outflow adopts a conical geometry.  The
solution with $\lambda_{0}^{\prime} =1.6$ in Fig.~\ref{fig1}
illustrates this regime. For even larger values of
$\lambda_{0}^{\prime}$, the solution switches to a global inflow
pattern ($\lambda_{0}^{\prime}=2.0$), with increasingly faster inflow
in the polar regions, which ultimately overcomes the inflow speed at
the equator ($\lambda_{0}^{\prime}=2.35$). At the end of this sequence
($\lambda_{0}^{\prime} \gta 2.35$), the flow approaches the
non-rotating, uniform density and uniform temperature limit: the 2D
solutions with conduction become Bondi-like, in a close analogy with
what we have seen for 1D solutions with conduction in \S3.  In this
limit, the value of $\rho \sim 5\dot m /(6\pi \alpha) =2.65$ (see
Appendix B). We note that along this entire sequence of solutions,
maximum inflow and outflow speed remained everywhere subsonic. As we
shall see below in more detail, most of these solution properties can
be understood in terms of the relative contributions of radial and
polar conduction terms to the flow energetic balance.

\begin{figure}
\plotone{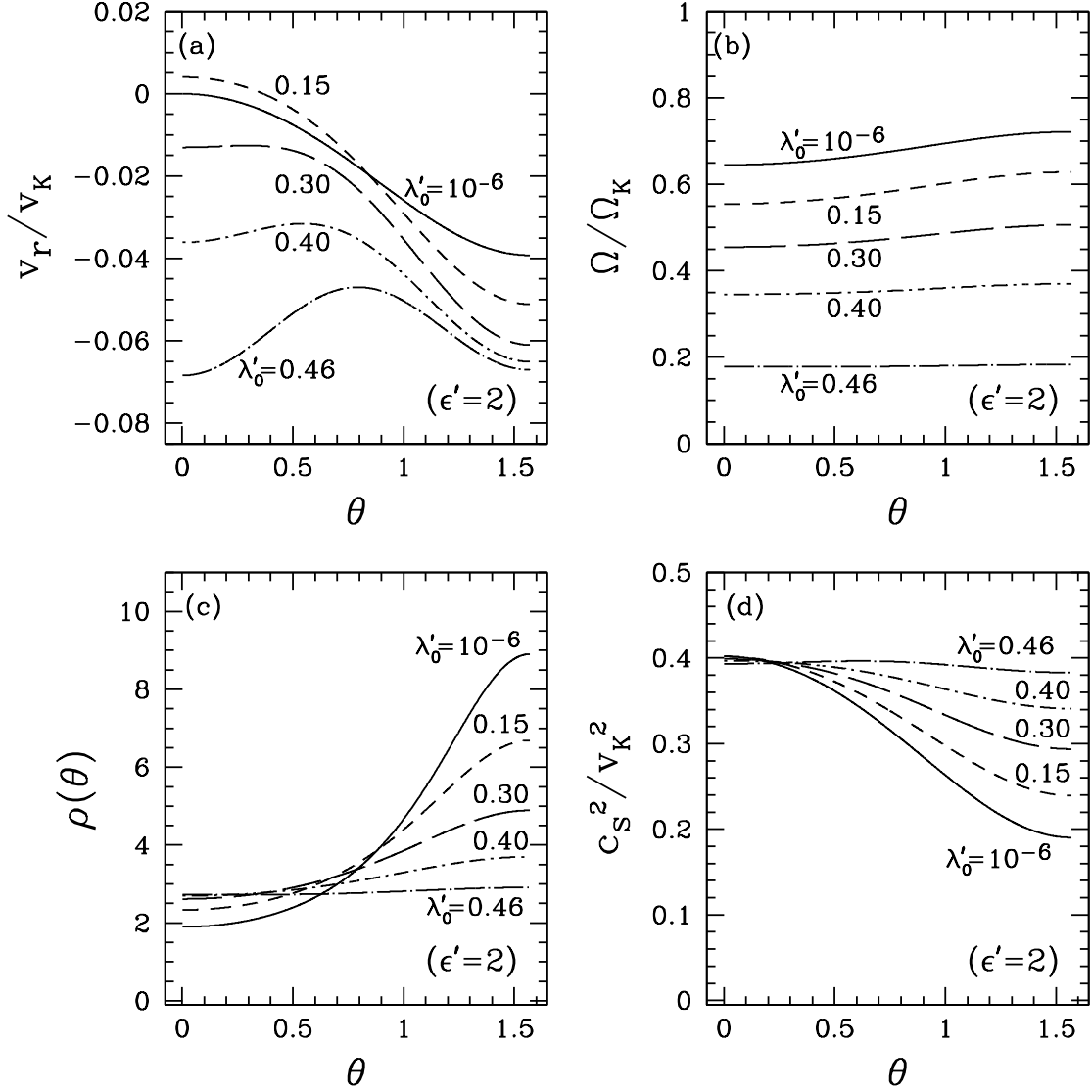}
\caption{Latitudinal profiles of flow dynamical quantities in five 2D
self-similar solutions with $\alpha=0.1$, $\epsilon^{\prime}=2$ and
varying levels of conduction (as measured by
$\lambda_{0}^{\prime}$). The notation is the same as in
Fig.~\ref{fig1}.}
\label{fig2}
\end{figure}

Figure~\ref{fig2} shows a comparable sequence of latitudinal profiles
in solutions with a reduced value of $\epsilon^{\prime}$ ($=2$) and
the same $\alpha=0.1$ as before. For small enough values of
$\lambda_{0}^{\prime}$ ($=10^{-6}$), the solutions are again close to
the original 2D ADAF solutions of Narayan \& Yi (1995a), with a more
spherically symmetric profile than in the $\epsilon^{\prime}=10$ case,
but still a $v_r$ value which is (nearly) zero at the pole.  As the
value of $\lambda_{0}^{\prime}$ is increased, a sequence similar to
that in Fig.~\ref{fig1} obtains, with a few important differences:
inflow and outflow velocities, outflow opening angles and the critical
$\lambda_{0}^{\prime}$ value at which the non-rotating limit is
reached are all smaller than in the $\epsilon^{\prime}=10$ case. In
addition, there is no conical outflow regime along the
$\lambda_{0}^{\prime}$ sequence in this subclass of solutions.

Figure~\ref{fig3} illustrates in a different way the flow geometry and
the relative importance of the rates of mass inflow and outflow, for
four of the solutions shown in Fig.~\ref{fig1} ($\alpha=0.1$,
$\epsilon^{\prime}=10$). Isodensity contours are shown by solid lines,
while color contours trace the flux of mass per unit solid angle
($=\rho v_r$). The scale of the cylindrical coordinates ($Rz$) used to
label the axes, as well as the isodensity contour levels, are
arbitrary because of the radially self-similar nature of the solutions
considered.

Figure~\ref{fig4} summarizes the most significant results for the
existence and the properties of outflows that emerged from our
systematic survey of the parameter space of hot accretion flow
solutions with conduction. Focusing on the specific $\alpha =0.1$
case, the various curves in Fig.~\ref{fig4} cover the
$(\epsilon^{\prime},\lambda_{0}^{\prime})$ space. The four panels show
(a) the flow radial velocity at the pole, $v_r(\theta = 0)$, in units
of the Keplerian value, (b) the angular velocity at the equator,
$\Omega(\theta = \pi/2)$, in units of the Keplerian value, (c) the
opening angle of outflowing regions, including conical outflow cases,
and (d) the ratio of outflowing to net inflowing mass rate (which is
also the accretion rate defined in Eq.~[\ref{eq:mdot}]).

A number of clear trends emerge from this parameter space survey.
Solutions with small $\epsilon^{\prime}$ values ($\lta 1$) do not
exhibit outflows, but only pure inflow regimes. Conversely, solutions
with the largest values of $\epsilon^{\prime}$ are the most efficient
at producing outflows. Furthermore, even among the solutions with
$\epsilon^{\prime} \gta 1$, there is only a finite range of
$\lambda_{0}^{\prime}$ values for which an outflow regime exists. The
larger $\epsilon^{\prime}$ is, the larger are the velocity, opening
angle and mass flux of the outflow, and so is the maximum value of
$\lambda_{0}^{\prime} / \epsilon^{\prime}$ beyond which the outflow
ceases to exist. We also find that solutions with conical outflows
exist only for sufficiently high $\epsilon^{\prime}$ values.  Finally,
for large $\epsilon^{\prime}$ solutions, the outflowing regime
persists even to very small values of $\lambda_{0}^{\prime} /
\epsilon^{\prime}$ but outflows become increasingly weaker, slower and
narrower. It should not be a surprise that these various flow
properties scale with the ratio $\lambda_{0}^{\prime} /
\epsilon^{\prime}$, given the form of the energy equation
(Eq.~[\ref{energy}]). All these trends are consistent with the
detailed results shown in Figs.~\ref{fig1} and~\ref{fig2}.

The increased efficiency of large $\epsilon^{\prime}$ solutions at
producing outflows can be interpreted as resulting from the larger
latitudinal temperature gradients present in these solutions. This
makes the contribution of the polar conduction term more important,
relative to the radial conduction term (which is tied to a fixed
radial temperature gradient set by self-similarity).  Mass outflow
rates $\sim 6\%$ of the net inflow rate, outflow velocities $\lta 2\%$
of the local Keplerian velocity and outflow opening angles $\sim
50$~degs are reached in solutions with $\epsilon^{\prime} =10$ and
$\lambda_{0}^{\prime} \simeq 0.1$. {For even larger values of
$\epsilon^{\prime}$, outflow properties continue to follow the trends
shown in Fig.~\ref{fig4}. For instance, we have found
$\epsilon^{\prime} =50$ solutions with mass outflow rates $\lta 13\%$
of the net inflow rate, outflow velocities $\lta 3\%$ of the local
Keplerian velocity and outflow opening angles $\lta
60$~degs. Numerical convergence becomes increasingly difficult at
large $\epsilon^{\prime}$ values, however, and we have not explored
this regime as systematically as the low $\epsilon^{\prime}$ one.}

Two additional features in Fig.~\ref{fig4} are worth emphasizing. All
solutions reach the non-rotating limit for values of
$\lambda_{0}^{\prime} / \epsilon^{\prime} \approx 0.24$ and $v_r \lta
-0.06~v_{\K}$. A derivation of these asymptotic values for the
non-rotating limit is provided in Appendix B.  We also find that there
is a wedge in the $\lambda_{0}^{\prime} / \epsilon^{\prime}$
vs. $\epsilon^{\prime}$ space, such that solutions do not exist for a
growing range of $\lambda_{0}^{\prime} / \epsilon^{\prime}$ as the
value of $\epsilon^{\prime}$ is increased. Despite careful efforts to
reach numerical convergence in these parts of the parameter space, we
have failed to find solutions of sufficient numerical accuracy (or at
all) in these regions. We have not clearly identified the origin of
this difficulty.  Let us stress that all the solutions shown in
figures, and otherwise discussed in this work, have reached
satisfactory numerical convergence.

\begin{figure}
\plotone{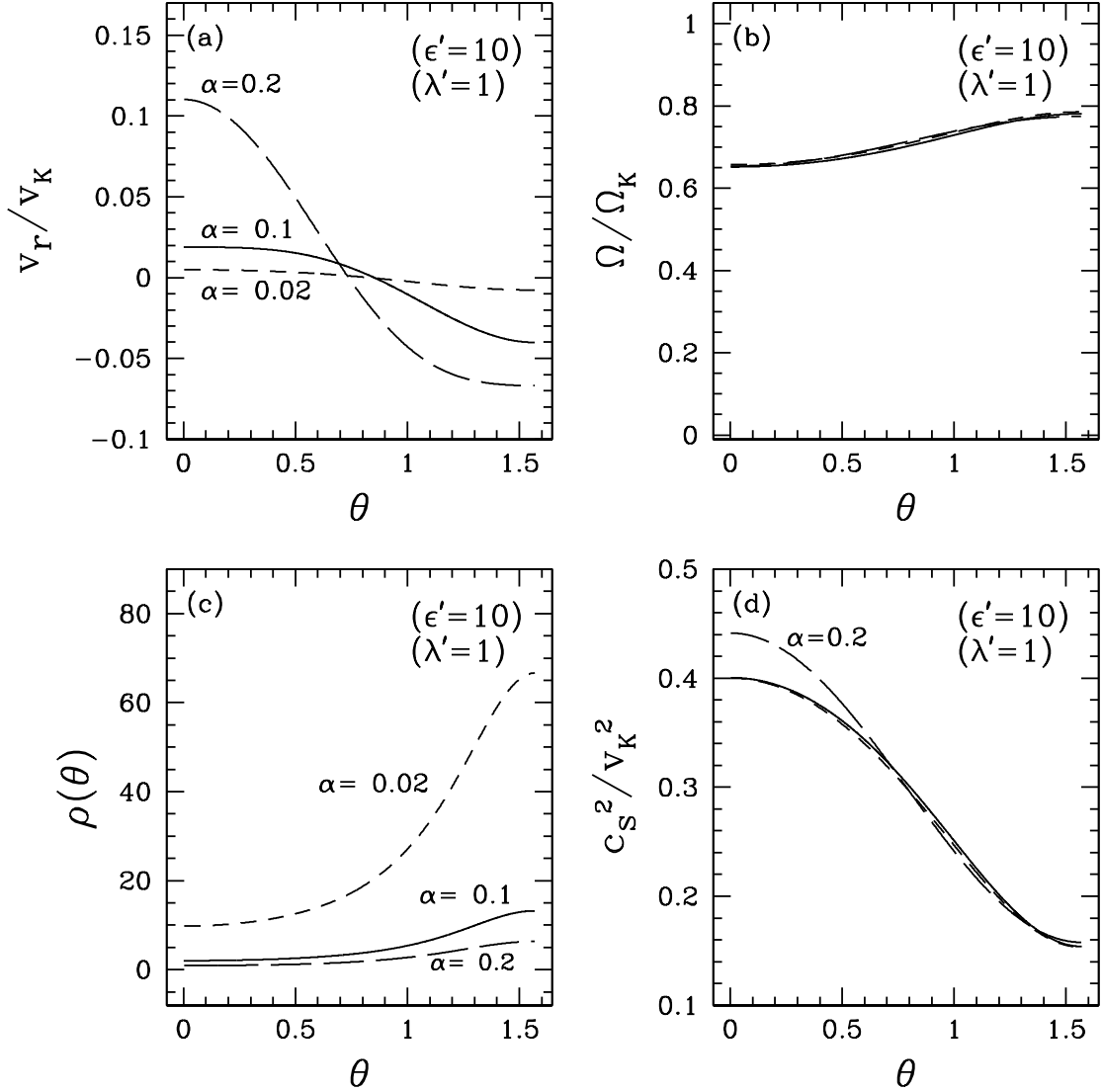}
\caption{Latitudinal profiles of flow dynamical quantities in 2D
self-similar solutions with $\epsilon^{\prime}=10$,
$\lambda_{0}^{\prime}=1$ and three different values of the viscosity
parameter $\alpha$: $0.02$ (short-dashed line), $0.1$ (solid line) and
$0.2$ (long-dashed line).}
\label{fig23}
\end{figure}

\begin{figure}
\plotone{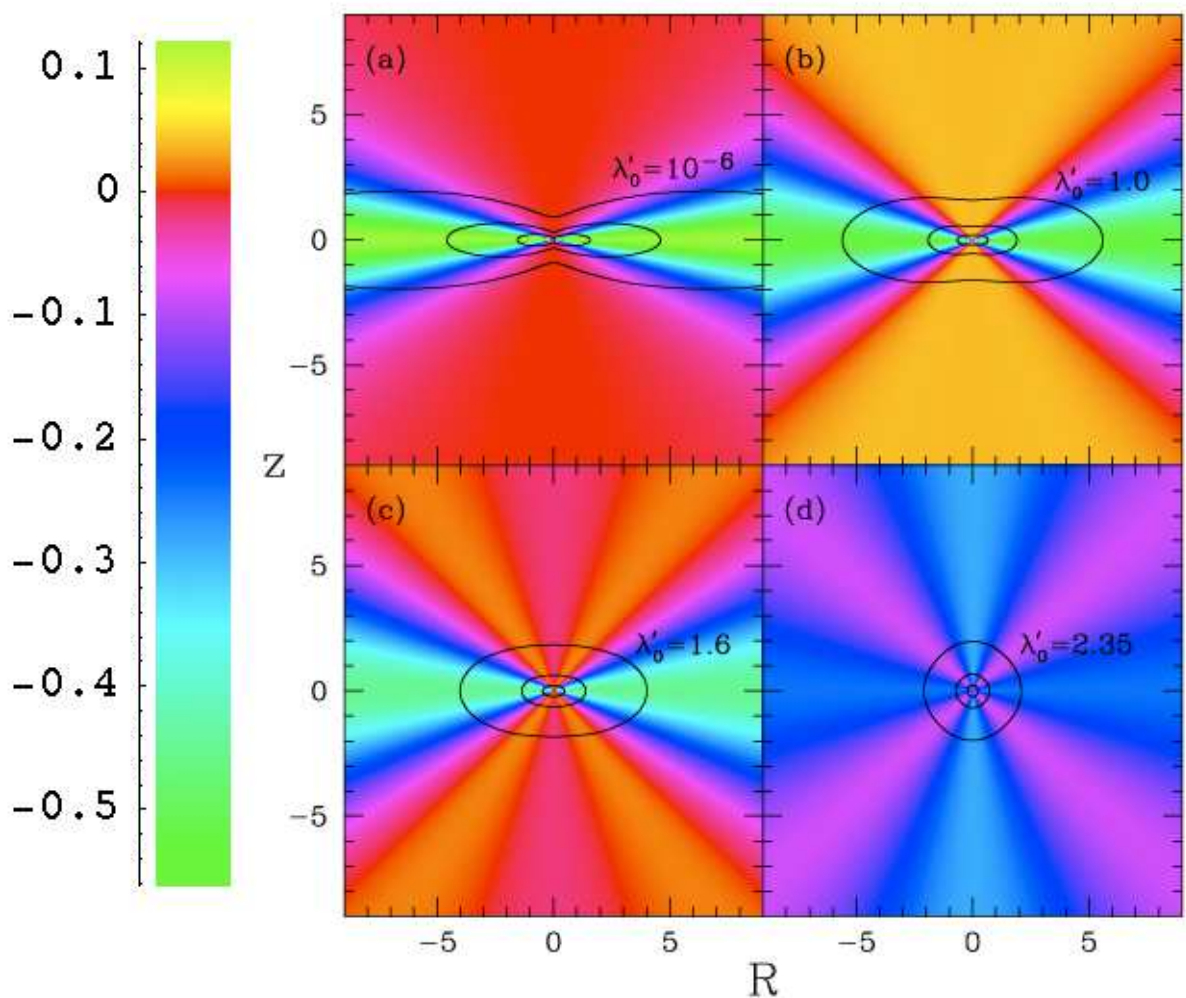}
\caption{Two-dimensional isodensity contours (black lines) and mass
flux contours (per unit solid angle; colors) for four of the solutions
shown in Fig.~\ref{fig1} with $\alpha=0.1$, $\epsilon^{\prime}=10$ and
$\lambda_{0}^{\prime}=10^{-6}$ (a), 1.0 (b), 1.6 (c), and 2.35
(d). The net inflow rate is normalized to $\dot m=1$ in all cases. The
axes are labeled in terms of cylindrical coordinates $R$ and $z$.}
\label{fig3}
\end{figure}

\begin{figure}
\plotone{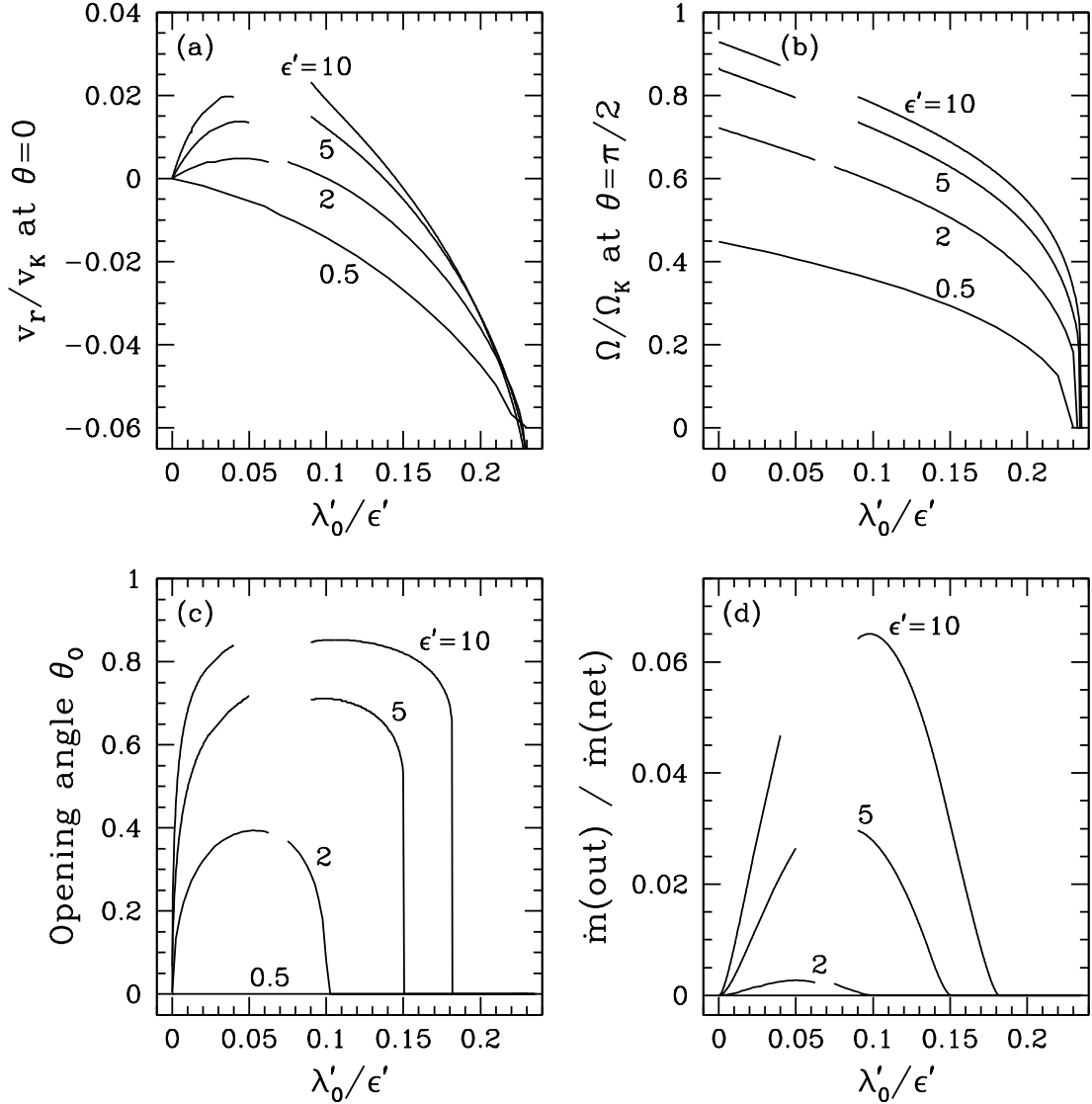}
\caption{Variations of several important flow properties as a function
of $\lambda_{0}^{\prime}/\epsilon^{\prime}$ in 2D solutions with
$\alpha=0.1$ and $\epsilon^{\prime}=0.5$, $2$, $5$ and $10$.  (a):
Flow velocity at the pole ($\theta=0$) in units of the Keplerian
value, $v_{\K}$. (b): Angular velocity at the equator ($\theta=\pi/2$)
in units of the Keplerian value, $\Omega_{\K}$.  (c): Opening angle of
the bipolar outflow, $\theta_{{\rm o}}$.  (d): Mass outflow rate,
$\dot m$(out), in units of the net accretion/inflow rate, $\dot
m$(net). Robust numerical solutions were not found within a narrow
wedge of $\lambda_{0}^{\prime}/\epsilon^{\prime}$ values. Solutions
cease to exist once the $\Omega=0$ limit is reached.}
\label{fig4}
\end{figure}

{We find that the value of the viscosity parameter, $\alpha$, affects
quantitatively (but not qualitatively) the properties of hot accretion
flow solutions with conduction. For larger values of $\alpha$, the
radial inflow and outflow speeds are increased and the overall density
scale is reduced, as expected from the original ADAF $\alpha$
scalings. Figure~\ref{fig23} shows a specific comparison of solutions
with $\epsilon^{\prime} =10$, $\lambda_{0}^{\prime} /
\epsilon^{\prime}=0.1$ and $\alpha = 0.02$, $0.1$ and $0.2$. The
outflow speed is much larger in the $\alpha = 0.2$ solution, reaching
a value $\simeq 0.11 \,v_{\rm K}$ at the pole. Yet, the mass outflow
rate ($\sim 9\% \, \dot m$) is not much greater in the $\alpha= 0.2$
solution than in the other two solutions ($\sim 6$-$7\% \, \dot m$)
because of the reduced density scale associated with faster radial
velocities. The outflow opening angles are comparable in all three
solutions. For consistency and conciseness, we focus on solutions with
$\alpha=0.1$ below.}

\subsubsection{Detailed Energy and Momentum Balance}

To understand better the various properties of these hot accretion
flows with conduction, let us now focus on the detailed energy and
momentum balance that they satisfy.

\begin{figure}
\plotone{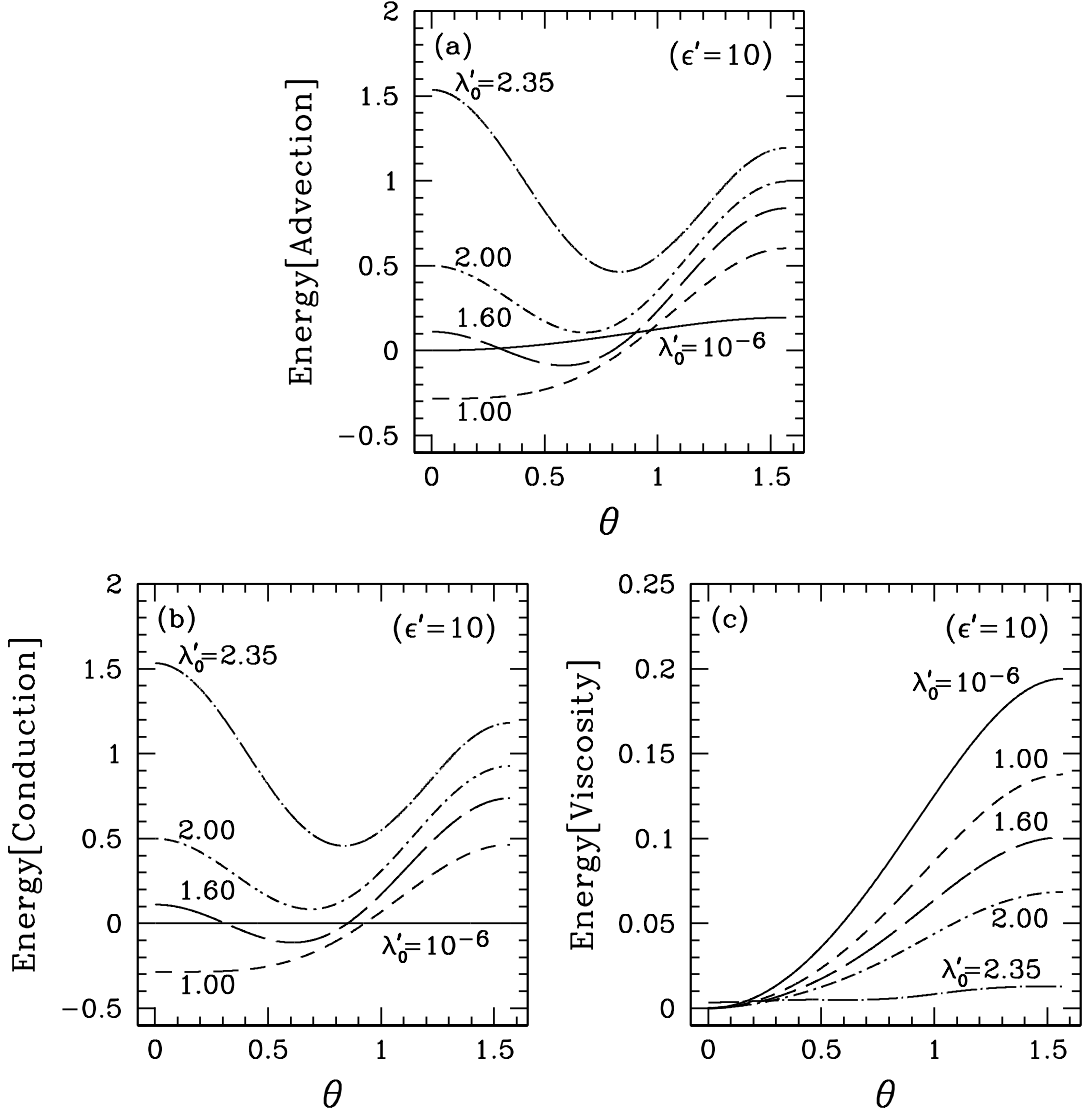}
\caption{Latitudinal profiles of the advection (a), conduction (b) and
viscous (c) terms in the 2D self-similar energy equation, for five
solutions with $\alpha=0.1$, $\epsilon^{\prime} =10$ and varying
degrees of conduction (as measured by $\lambda_{0}^{\prime}$). Note
the different vertical scales.}
\label{fig5}
\end{figure}

Figure~\ref{fig5} shows latitudinal profiles of the advection term
(a), on the LHS of the energy equation (Eq.~[\ref{energy}]), and of
the conduction (b) and viscous (c) terms, on the RHS of
Eq.~(\ref{energy}), for the same specific sequence of solutions as
shown in Fig.~\ref{fig1} ($\alpha=0.1$, $\epsilon^{\prime} =10$). In
the solution with negligibly small conduction ($\lambda_{0}^{\prime}
=10^{-6}$), the advection and the viscous terms balance each other
exactly and heat advection acts to cool the flow (which corresponds to
positive values in panel a). In the solution with
$\lambda_{0}^{\prime} =1.0$, however, the magnitude of the conduction
term is already very important, being comparable to or larger than the
magnitude of the viscous term in most of the flow. In addition,
conduction both cools the flow in the polar regions (which corresponds
to negative values in panel b) and heats it up in the equatorial
regions. As a result of this cooling of the polar regions (where the
contribution from the viscous term is comparatively small, simply by
geometry), the balance between these various terms permits a negative
advection term in the polar regions, which corresponds to advective
heating and outflow.

As the value of $\lambda_{0}^{\prime}$ reaches $1.6$, the latitudinal
profile of temperature starts to flatten significantly (see
Fig.~\ref{fig1}), thus reducing the contribution of the polar
component of conduction. The radial component, on the other hand, is
strictly positive and can only increase in magnitude with the value of
$\lambda_{0}^{\prime}$ (because of the radial self-similarity of the
solutions). As a result, the net conduction term switches sign again
in the polar regions (which are now heated by conduction), but it is
able to remain somewhat negative at intermediate values of the polar
angle, $\theta$. This situation leads to the conical outflow regime
previously identified.

Finally, for even larger values of $\lambda_{0}^{\prime} (\geq 2)$,
the radial component of conduction dominates everywhere over the polar
component (which is further reduced by the continued flattening of the
temperature profile; Fig.~\ref{fig1}), thus forcing the net conduction
term to act strictly as a heat source (as in the 1D solutions of \S3).
The energy equation is then satisfied only if advection acts to cool
the flow everywhere, which must then be globally inflowing. For the
largest values of $\lambda_{0}^{\prime}$ considered, advection very
nearly balances conduction as the viscous term drops precipitously and
the non-rotating limit is reached. In good analogy with what we have
seen for 1D solutions with conduction in \S3, the non-rotating
Bondi-like limit corresponds to strict equality between conduction and
advection.

From this examination of energy balance in our radially self-similar
solutions, it is clear that the key to thermal outflows is the
existence of strong enough cooling in the flow for advection to be
allowed to act as a heat source. This is preferentially achieved in
polar regions, simply because these are regions where the rate of
viscous dissipation must drop by geometrical constraint.

Let us comment here on the fact that the advection parameter, $f$, is
actually ill-defined in our solutions with conduction. Since $f$ is
associated with the viscous dissipation term in Eq.~(\ref{energy}), it
represents a form of cooling which is proportional to the rate of
viscous dissipation. As a result, in solutions with an energy balance
dominated by conduction (e.g. solutions with large
$\lambda_{0}^{\prime}$ in Fig.~\ref{fig5}), a small value of $f$ ($\ll
1$) does not correspond to a globally efficient cooling anymore, as it
used to in the original ADAF solutions. This undesirable situation
would disappear in 2-temperature solutions with more realistic
descriptions of cooling, such as the ones described by Narayan \& Yi
(1995b). Note also that this is not a critical flaw for our analysis:
the energetics of all the solutions discussed here depends only on the
value of $\epsilon^{\prime} \equiv \epsilon/f$, so that large
$\epsilon^{\prime}$ values can equally well be interpreted as
corresponding to low $\gamma$ values.

\begin{figure}
\plotone{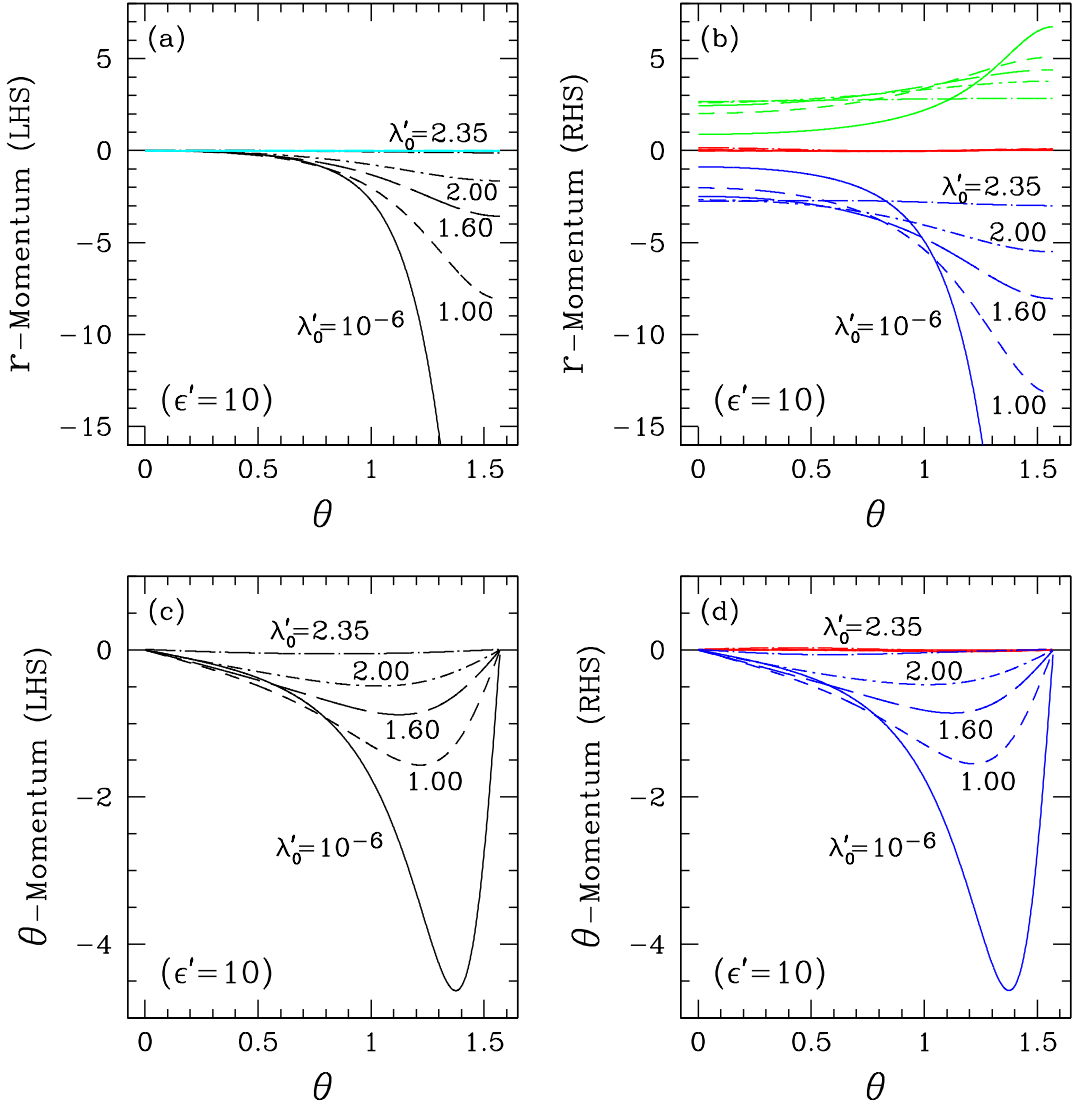}
\caption{Latitudinal profiles of terms in the $r$- and
$\theta$-components of the momentum balance equation in five solutions
with $\alpha=0.1$, $\epsilon^{\prime} =10$ and varying degrees of
conduction (as measured by $\lambda_{0}^{\prime}$). Top panels: left-
and right-hand sides of the radial momentum equation
(Eq.~[\ref{moma}]).  (a) shows the magnitude of the inertial term
(cyan) and the centrifugal term (black) on the LHS of the $r$-momentum
equation.  (b) shows contributions to the RHS of this same equation:
gravitational (blue), pressure gradient (green) and viscous (red)
terms.  Bottom panels: left- and right-hand sides of the latitudinal
momentum equation (Eq.~[\ref{momb}]).  (c) shows the magnitude of the
centrifugal term (black).  (d) shows contributions from the pressure
gradient (blue) and viscous (red) terms.}
\label{fig6}
\end{figure}

For completeness, we now examine in detail the momentum balance in
these same solutions ($\alpha=0.1$, $\epsilon^{\prime}
=10$). Figure~\ref{fig6} shows latitudinal profiles of various terms
in the radial and polar components of the momentum equation
(Eqs.~[\ref{moma}] and~[\ref{momb}], respectively). The magnitude of
the various terms on the LHS and the RHS of these equations are
labeled with different colors in the panels of Fig.~\ref{fig6}, for
the same sequence of increasing $\lambda_{0}^{\prime}$ values as
before.  Note that we ignore the azimuthal momentum equation, which
simply expresses angular momentum conservation.

For a negligibly small level of conduction ($\lambda_{0}^{\prime}
=10^{-6}$), the $r$-momentum balance is dominated by pressure support
against gravity in the polar regions, while rotational support becomes
gradually dominant in the equatorial regions. As the value of
$\lambda_{0}^{\prime}$ is increased, the flow becomes hotter, more
spherically symmetric and the $r$-momentum balance tends to be more
dominated by pressure support at all values of the polar angle,
$\theta$. The inertial and viscous terms are never very important for
$r$-momentum balance along this $\lambda_{0}^{\prime}$ sequence. As
for the $\theta$-momentum equation, for all values of
$\lambda_{0}^{\prime}$ considered, it is satisfied through a balance
between latitudinal pressure gradients and projected centrifugal
force, with little contribution from the viscous terms. Perhaps one of
the most interesting property of the profiles shown in Fig.~\ref{fig6}
is that they show explicitly that there is no distinction, from the
point of view of momentum balance, between inflowing and outflowing
regions of the flow. This confirms that outflows, when they exist, are
purely determined by energetic considerations. This further justifies
us classifying them as ``thermal'' in origin.

\subsubsection{The Bernoulli Parameter} \label{sec:bern}

Much of the discussion in the literature on the likelihood of outflows
from hot accretion flows is articulated around the value of the
Bernoulli parameter, $B_{\rm e}$ (or, more generally, the Bernoulli
function). Narayan \& Yi (1994; 1995a) first emphasized the positivity
of $B_{\rm e}$ in their solutions and raised the possibility that this
would lead to bipolar outflows. Furthermore, Blandford \& Begelman
(1999)'s motivation and method to develop ADIOS is rooted in the
notion that hot accretion flows with positive $B_{\rm e}$ must have
outflows and that their structure and properties should be accordingly
modified, leading to a mass accretion rate varying with radius. The
relevance of the value of $B_{\rm e}$ for the generation of outflows
has also been discussed in the context of numerical simulations
(e.g. Stone et al. 1999) and it has been put into question
(e.g. Abramowicz et al. 2000).  Since our hot accretion flow solutions
with conduction produce outflows spontaneously, it is of considerable
interest to study the relation between these outflows and the value of
$B_{\rm e}$ in the solutions.

\begin{figure}
\plotone{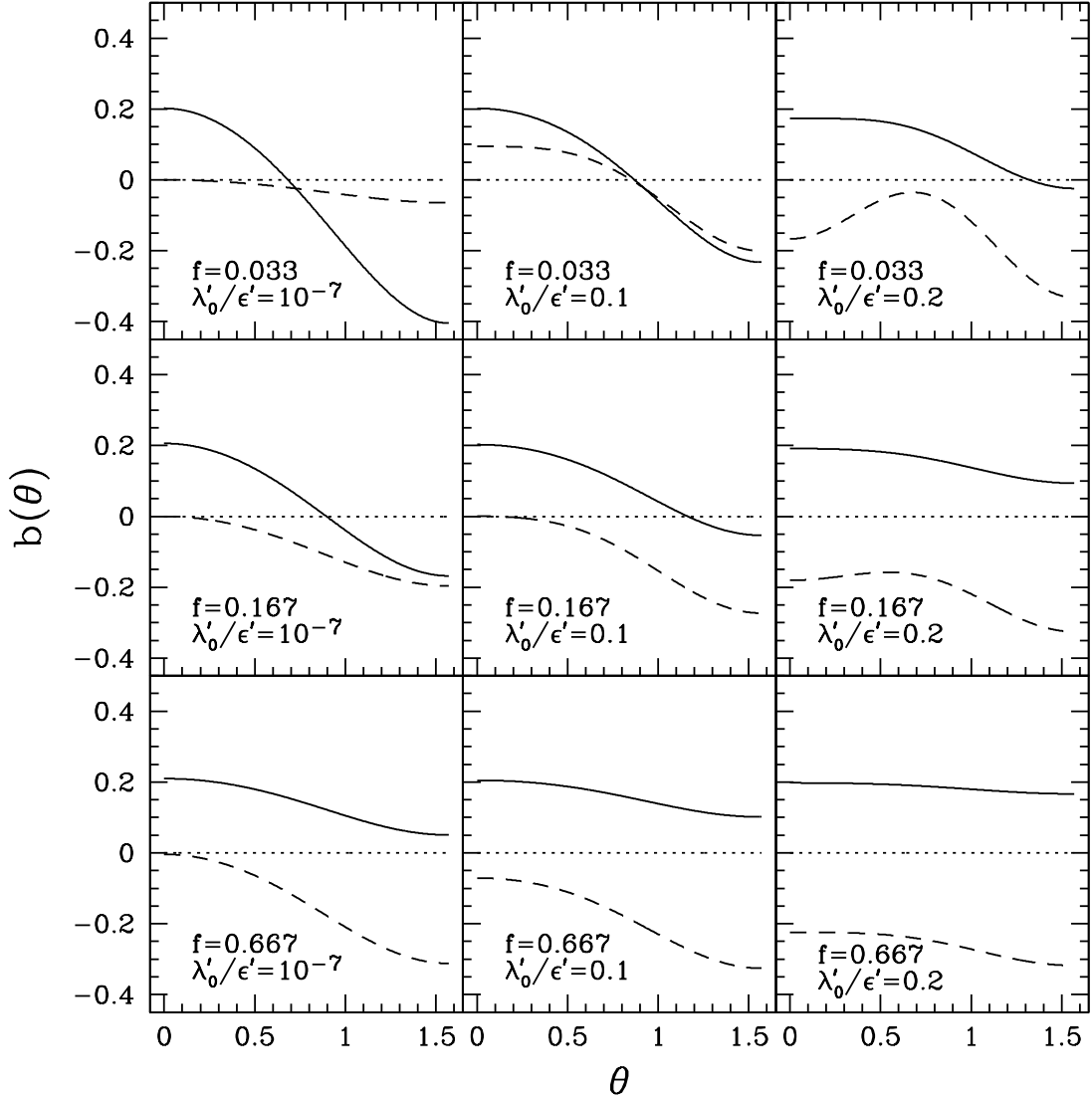} 
\caption{Latitudinal profiles of the dimensionless Bernoulli
parameter, $b(\theta)$ (solid lines), and of the radial velocity,
$v_r(\theta)/v_{\K}$ (multiplied by an arbitrary factor 5; dashed
lines) in 2D self-similar solutions with $\gamma=1.5$.  The various
panels compare these profiles for different values of the advection
parameter, $f$, and varying degrees of conduction (as measured by the
scaled parameter $\lambda_{0}^{\prime}/\epsilon^{\prime}$).
Horizontal dotted lines were added in each panel to help identify
inflowing vs. outflowing regions. Note that the three rows of panels
correspond to solutions with $\epsilon^{\prime} \simeq 10$ (top), $2$
(middle) and $0.5$ (bottom).}
\label{fig7}
\end{figure}

Figure~\ref{fig7} shows latitudinal profiles of the dimensionless
Bernoulli parameter, $b(\theta)= B_{\rm e}/ (r \Omega_{\rm K})^2 $
(same notation as Narayan \& Yi 1995a) in a set of solutions with
$\alpha =0.1$, $\gamma=1.5$ and different values of $f$ and
$\lambda_{0}^{\prime}/\epsilon^{\prime}$.  In addition to $b(\theta)$,
scaled latitudinal profiles of the radial velocity, $v_r$, are shown,
for an easy identification of inflowing vs. outflowing regions. The
nine panels in Fig.~\ref{fig7} are organized in such a way that the
three rows correspond to $\epsilon^{\prime} \simeq 10$, $2$ and $0.5$,
from top to bottom, with increasing values of
$\lambda_{0}^{\prime}/\epsilon^{\prime}$ from left to right.

\begin{figure}
\plotone{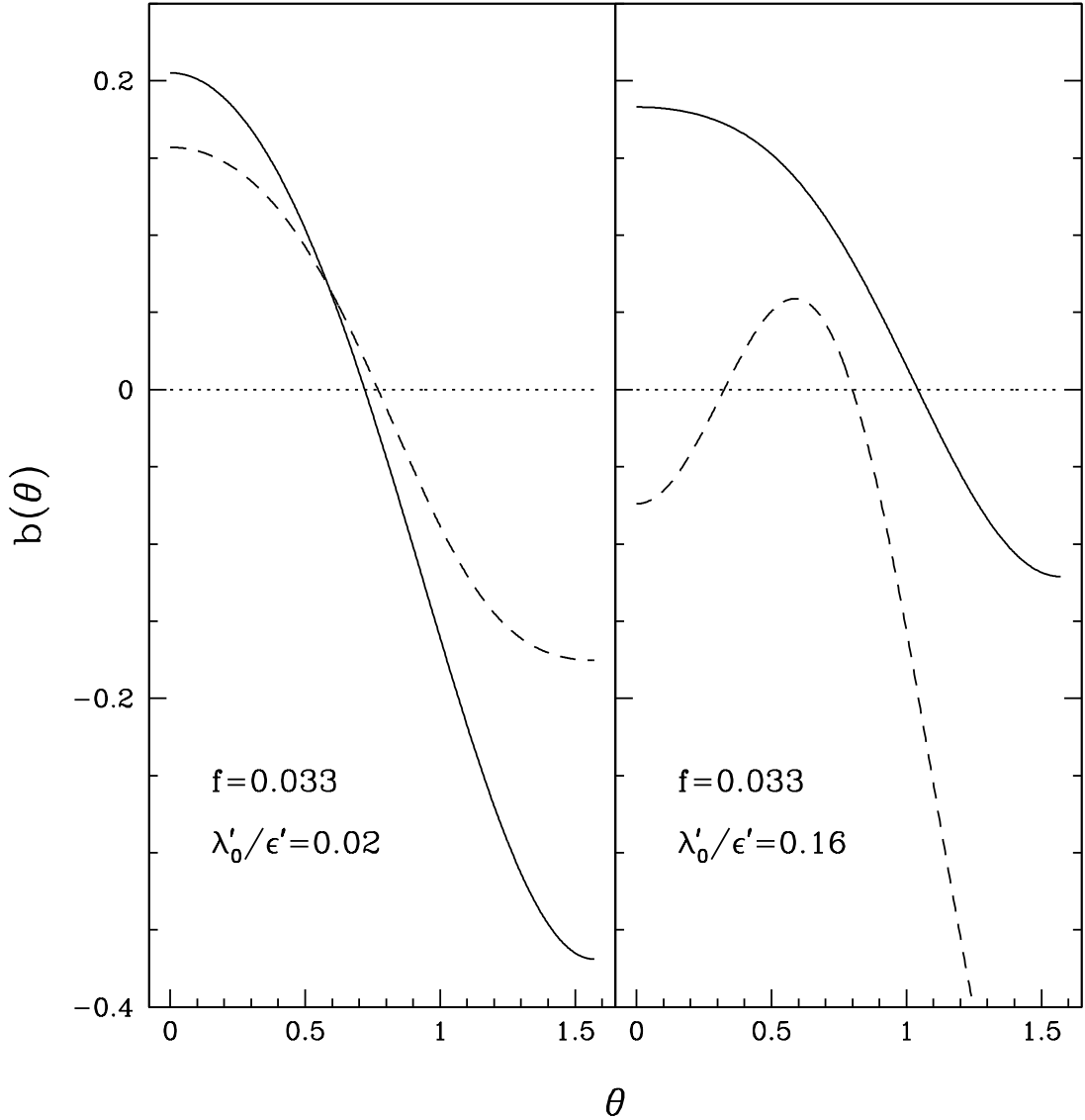} 
\caption{Latitudinal profiles of the dimensionless Bernoulli
parameter, $b(\theta)$ (solid lines), and of the radial velocity,
$v_r(\theta)/v_{\K}$ (multiplied by an arbitrary factor 5; dashed
lines) in 2D self-similar solutions with $\gamma=1.5$, $f=0.033$
(hence $\epsilon^{\prime}\simeq 10 $) and two values of the scaled
conduction parameter, $\lambda_{0}^{\prime}/\epsilon^{\prime}$.  The
two panels show solutions of special interest, with a conical outflow
geometry (right panel) and with a marginal outflowing region having
$b(\theta) < 0$ (left panel).  Horizontal dotted lines were added in
each panel to help identify inflowing vs. outflowing regions. }
\label{fig8}
\end{figure}

From this ensemble of solutions, one concludes that the positivity of
$b(\theta)$ does not in general discriminate between inflowing and
outflowing regions in our solutions. {Although the solution
corresponding to $f=0.033$ and $\lambda_{0}^{\prime}/\epsilon^{\prime}
= 0.1$ would seem to indicate it does, this is in fact a coincidence
since we find many counterexamples. Notice in particular in
Fig.~\ref{fig7} how solutions with $b(\theta)$ changing from positive
at the pole to negative at the equator can be alternatively (nearly)
static, outflowing or inflowing at the pole. Similarly, solutions with
globally positive values of $b(\theta)$ can be globally inflowing
without any difficulty.

The additional panels in Figure~\ref{fig8} show two particularly
striking examples of the loose relation between the sign of
$b(\theta)$ and the inflow/outflow properties of our solutions. In one
solution, with $\lambda_{0}^{\prime}/\epsilon^{\prime} = 0.16$, the
outflow has a conical geometry. While the value of $b(\theta)$
decreases monotically with $\theta$, the radial velocity is clearly
not monotonic, with two sign changes. In the other solution, with
$\lambda_{0}^{\prime}/\epsilon^{\prime} = 0.02$, the outflow is polar
but there is a small regions at mid-latitudes where outflowing gas has
negative $b(\theta)$ value. Although this type of ``violation'' is
generally marginal, it clearly demonstrates that there is no strict
relation between the sign of $b(\theta)$ and the inflow/outflow
properties of gas in our solutions.  It should be noted that for large
enough values of $\lambda_{0}^{\prime}$, one would expect the
Bernoulli parameter (which is an adiabatic, inviscid quantity) to lose
its physical significance, since the flow must then become strongly
diabatic. We suspect that the $b(\theta) < 0$ outflow region shown in
the left panel of Fig.~\ref{fig8} and the globally inflowing solutions
with globally positive $b(\theta)$ values shown in Fig.~\ref{fig7} are
all manifestations of this loss of significance of $b(\theta)$ in
flows with significant conduction.}

\section{Relevance to Observational Trends}

It is difficult to evaluate, from self-similar solutions alone, what
the radiative efficiency of hot accretion flows with conduction may
be. In our solutions, the flow structure adjusts so that the
divergence of the conduction flux generated in the innermost regions
is balanced by an increased level of entropy advection. There will be
no extra source of energy available in the presence of conduction (in
fact less is available in our solutions, since the viscous dissipation
rate drops with $\Omega$), but the dissipated energy is redistributed
differently in flows with conduction than in flows without. This
redistribution affects the local flow properties (density,
temperature), which in turn affect its cooling properties.

Using our 1D solution scalings (\S3) for simplicity, it is possible to
estimate crudely the modified level of emission expected relative to a
standard ADAF. Free-free emission scales as $\rho^2 T^{1/2} \propto
(c_{s0} v_0^2)^{-1}$ for a given value of the accretion rate, $\dot
M$.  We find that this emission is effectively reduced, e.g. by as
much as $\sim 25\%$ in the 1D solution of Fig.~\ref{fig:one} with
$\gamma =1.5$ (and much more in the $\gamma =1.1$ solution). Even
though conduction heats up the gas locally, the reduced density
resulting from the larger inflow speed dominates, leading to a net
decrease in the expected level of free-free emission. On the other
hand, the very steep dependence of synchrotron emission on the
electron temperature (e.g. Mahadevan 1997) suggests that hotter
solutions (with conduction) may be radiatively more efficient from the
point of view of that process. These are only rough scalings,
however. In light of several complications expected in global
solutions with conduction (see discussion in \S6), it is unlikely that
we can determine, from self-similar solutions alone, how the global
radiative efficiency of a hot accretion flow may be affected by
conduction. We postpone a thorough discussion of this important issue
to future work and focus here instead on the possible relevance of the
outflows found in our solutions to a variety of observationally
established trends for accreting black holes.

In recent years, much progress has been made in understanding the
outflow/jet properties of accretion flows around black holes in X-ray
binaries and Active Galactic Nuclei. Various observational studies
suggest that, in both classes of systems, there is a regime of steady
outflow when black holes accrete well below the Eddington rate (for
luminosities $\lta 0.01-0.1 L_{\rm Edd}$), while the outflow regime
becomes flaring/unsteady at higher accretion rates (e.g. Ho 2002;
Gallo, Fender, Pooley 2003; Merloni, Heinz \& Di Matteo 2003; Fender,
Belloni \& Gallo 2004; Nipoti, Blundel \& Binney 2005; Sikora, Stawarz
\& Lasota 2006). {Some of the latest results support the notion that
black holes accreting at sub-Eddington rates do so through an
advective form of accretion (e.g. Greene, Ho \& Ulvestad 2005;
Koerding, Fender \& Migliari 2006), but there is some evidence in the
``low/hard'' state of at least two Galactic black hole systems (Miller
et al. 2006) that a standard thin disk remains embedded in the hot
flow (a situation which may favor outflows according to our large
$\epsilon^{\prime}$ solutions). It is tempting to relate these various
observational trends with our results on thermal outflows from steady,
hot accretion flows with conduction.}

The association would be strengthened if one could show that thermal
conduction is expected to be equally important in hot accretion flows
around stellar-mass and supermassive black holes. After all, we made
the case for weakly-collisional accretion conditions only for systems
harboring supermassive black holes in \S2. While the temperature
profile does not depend on black hole mass in self-similar solutions
scaled in units of the Schwarzschild radius, density profiles do.  To
address this question, we compare the scalings for the coefficient of
turbulent viscosity ($\nu = \alpha c_s^2 / \Omega_{\rm K}$) and the
Spitzer coefficient of thermal diffusivity ($\chi \propto
T^{5/2}/\rho$), using for simplicity the ADAF scalings in
Schwarzschild units of Narayan et al. (1998b).  The comparison of these
two diffusivity coefficients is the relevant one since it captures the
relative magnitude of ``viscous'' and conduction effects on any scale
of interest. We find that the two coefficients scale identically
(linearly) with the mass of the central black hole, which suggests
that conduction would indeed be equally important in hot accretion
flows around stellar-mass and supermassive black holes. This
conclusion is also consistent with the similar scalings with black
hole mass of the accretion time and collision times discussed by
Mahadevan \& Quataert (1997).  It would be interesting to strengthen
this comparison by using, as much as possible, observational
constraints on hot accretion flows in X-ray binaries, as we have done
for AGN in \S2.

Establishing on a firmer ground the relevance of the thermal outflows
found in our solutions to observational trends for accreting black
holes will clearly require additional work. In the meantime, we note
that our solutions allow a direct scaling of the expected outflow
power, per unit accreted mass, which is a quantity of obvious
interest. Recalling that the Bernoulli constant, $b(\theta)$,
represents the asymptotic value of the energy of gas escaping from the
central mass gravity, we can use it as a measure of the outflowing
energy rate and compare that to the rate of energy release due to
accretion. The outflow power is simply $P_{\rm out} = \dot M_{\rm out}
b(\theta) v_{\rm K}^2$. The accretion power is $P_{\rm acc} \simeq 0.1
\dot M_{\rm in} c^2$ , if we pick a radius of maximum energy release
of $\sim 3$ Schwarzschild radii to fix the accretion efficiency to
$\sim 0.1$ as is usually done. At $3 R_{\rm s}$, $v_{\rm K}^2 \sim c^2
/ 6$. For a ratio of $\dot M_{\rm out}/ \dot M_{\rm in} \sim 0.05$
(solutions with $\epsilon^{\prime} = 10$ in Fig.~\ref{fig4}) and a
corresponding value of $b(\theta) \sim 0.15$ for outflowing regions
(Fig.~\ref{fig7}), we deduce a ratio $P_{\rm out} / P_{\rm acc}$ that
reaches up to $ \sim 1 \%$ in our solutions. Therefore, provided that
solutions with realistic cooling properties can be constructed with
radiative efficiencies $< 10^{-3}$ (i.e. less than a hundredth of the
fiducial $0.1$ value), it may be possible to explain the ``jet-power
dominance'' of radiatively-inefficient accretion flows which has been
recently emphasized by several authors (e.g. Nagar, Falcke \& Wilson
2005; Koerding et al. 2006; Allen et al. 2006). Interestingly, values
of $P_{\rm out} / P_{\rm acc} \lta 1\%$ as we just derived are
comparable to the values of $P_{\rm jet} / P_{\rm Bondi}$ of a few
percents inferred by Allen et al. (2006).

In this context, understanding how thermal outflows may relate to the
relativistic jets of accreting black holes will likely be
important. After all, outflow velocities decrease with radius as
$r^{-1/2}$ in our solutions (by virtue of self-similarity), which is a
rather unappealing feature. Thermal outflows could be collimated and
accelerated by their environment or, alternatively, they could
contribute to the collimation and acceleration of narrower and faster
jets, such as the ones emerging in relativistic MHD numerical
simulations of adiabatic tori (e.g. DeVilliers et al. 2005; Hawley \&
Krolik 2006).

One of the best systems available to us for the study of
radiatively-inefficient accretion flows (and related outflows) is
Sgr~A*. In recent years, the focus of theoretical work to interpret
the rich set of data has in fact shifted to include a jet or outflow
component (e.g. Falcke \& Biermann 1999; Falcke \& Markoff 2000; Melia
\& Falcke 2001; Markoff et al. 2001; Yuan, Markoff \& Falcke
2002). Observations of the size (e.g. Bower et al. 2004; Shen et
al. 2005; Yuan, Shen, Huang 2006), variability (e.g. Yuan, Quataert \&
Narayan 2003; Yusef-Zadeh et al. 2006) and polarization (e.g. Agol
2000; Quataert \& Gruzinov 2000b; Bower et al. 2003; Marrone et
al. 2006) of this source all have tremendous potential to further
constrain the properties of the hot accretion flow and its outflow,
and they may be able to help us discriminate between the thermal
outflows seen in our solutions and other outflow alternatives. Direct
constraints on outflow geometries for accreting black holes are
generally scarce, but it is interesting to note that Junor, Biretta \&
Livio (1999) observe an opening angle $\sim 60$~degs at the base of
the radio jet in M87, in line with the wide outflow geometries found
in our solutions. On the other hand, other geometries have been
proposed for other classes of systems (e.g. Elvis 2000).

The motivation to understand the nature and properties of outflows
from accreting black holes goes beyond what we have just outlined, as
outflows may be an efficient process by which black holes influence
their environments. Recently, this general point has been raised in
the specific context of radiatively-inefficient accretion flows by
several groups. Soria et al. (2006a,b; see also Di Matteo, Carilli \&
Fabian 2001) invoked slow outflows to effect a feedback on the rate of
Bondi accretion in several low-luminosity AGN. The wide outflow
geometries in our solutions may suggest that the conceptual use of a
simple Bondi spherical inflow geometry to estimate gas capture rates
is inappropriate if indeed a large fraction of the black hole capture
sphere is the site of gas outflow.  The work of Nagar et al.  (2005)
and Allen et al. (2006) further suggests that feedback due to outflows
associated with radiatively-inefficient accretion flows can reach far
beyond the black hole immediate vicinity and affect star formation in
the host galaxy, or even the surrounding hot gas when the host is
located at the center of a group or a cluster of galaxies. It will be
interesting to develop these arguments, and more detailed feedback
scenarios (e.g. Vernaleo \& Reynolds 2006), in light of what we have
learned in our study of thermal outflows.

\section{Discussion and Conclusion}

The possibility that tangled magnetic fields strongly limit the
efficiency of thermal conduction in a hot flow constitutes a major
theoretical uncertainty for our models. In recent years, a similar
issue has been discussed in the context of galaxy cluster cooling
flows. Several studies have concluded that magnetic fields may not
limit too strongly the efficiency of thermal conduction (e.g. Narayan
\& Medvedev 2001; Gruzinov 2002; see also Chandran \& Cowley 1998;
Chandran et al. 1999), but this remains somewhat of an open issue. If
tangled magnetic fields were to limit the effective mean free paths in
the hot flow to $l_{\rm eff} \lta R$, conduction would likely still
have important consequences for the flow structure and properties. If,
on the other hand, tangled magnetic fields act to make $l_{\rm eff}
\ll R$, a much weaker role for conduction (if any) is to be expected.
Explicit numerical simulations with anisotropic conduction along field
lines in an MHD turbulent medium could greatly help in settling this
issue (see, e.g., Cho \& Lazarian 2004; Sharma et al. 2006). It should
be emphasized that the energy transport mechanism which permits the
existence of outflows in our solutions does not have to be thermal
conduction. Any form of heat transport (e.g. turbulent, as emphasized
by Balbus 2004) acting like thermal conduction would have similar
consequences for the flow.

One of several limitations of our self-similar hot accretion flow
solutions with conduction is their 1-temperature structure. Narayan \&
Yi (1995b) have shown how crucial it is to account for the
2-temperature structure of the hot flow to obtain solutions with
realistic cooling properties. In such solutions, the expected
decoupling of electron and ion temperatures in the inner regions of
the hot flow would modify the role of conduction. For one thing, ions
may be partially insulated from electron conduction. In addition,
self-similarity is broken in the 2-temperature regime and the electron
temperature profile flattens to a sub-virial slope (e.g. Narayan et
al.  1998b). In global solutions, conduction is expected to flatten
both the polar and the radial temperature profiles (while it only
affects the polar profile in the radially self-similar solutions
discussed here). Furthermore, if a saturated form of conduction is
appropriate, the saturated flux should be reduced in regions of the
flow where the electrons become relativistic, since their ``free
streaming'' velocity then becomes limited by the speed of
light. Finally, the break-down of self-similarity implies that the
radial contribution to conduction, which acts only to heat up the flow
in our solutions, will have to be done at the expense of inner regions
that will now be cooling by conduction.  Clearly, it will be
interesting to see how all these effects combine with each other in
global models of hot accretion flows with conduction.

A potentially more fundamental limitation of our work may be the use
of simple fluid equations to describe a weakly-collisional plasma
obeying a more complex set of equations.  For instance, Balbus (2001)
has argued that the dynamical structure of hot flows could be affected
by the anisotropic character of conduction in the presence of magnetic
fields (see also Parrish \& Stone 2005), while Quataert et al. (2002)
have emphasized additional effects due to the anisotropic pressure
tensor present in the weakly-collisional regime (see also Sharma et
al. 2006). On the other hand, it is possible that a scale analysis
could justify the use of simpler fluid equations to describe the
dynamics of the weakly-collisional magnetized plasma on the largest
scales of interest in the context of hot accretion flows. Such an
analysis would clearly be valuable since essentially all the
literature on hot accretion flows relies on collisional fluid
equations (magnetized or not).

In recent years, much emphasis has been put on reducing the rate at
which gas is accreted by a central black hole, within a hot flow,
either by arguing that a negative Bernoulli constant is required for
an accretion solution with imposed winds to be physically viable
(Blandford \& Begelman 1999) or by invoking a gradual reduction in the
rate of mass accretion in the flow due to the effects of convection
(Narayan et al. 2000; Quataert \& Gruzinov 2000a). Our results,
combined with the possibility that conduction may be important in hot
accretion flows, could pose a challenge to both of these scenarios. We
have already seen that the sign of the Bernoulli constant does not
determine the existence of thermal outflows, or even discriminate
between inflowing and outflowing regions, in our solutions with
conduction. Our solutions also show that a radially varying $\dot m$
is not necessarily associated with the existence of such outflows.  In
addition, if conduction is important on all scales of interest in the
hot flow, convection may be suppressed altogether when a displaced
fluid element efficiently reaches thermal equilibrium with its
environment through conduction.

Obviously, our results do not exclude the existence of other classes
of spontaneously outflowing solutions with radially varying $\dot m$,
density profiles deviating from the canonical $r^{-3/2}$, more
realistic flow patterns allowing for non-zero latitudinal velocities,
polar funnels, or a role for other radial and latitudinal
energy/momentum transport mechanisms (see, e.g., the issues of
adiabatic stability discussed in Blandford \& Begelman 2004, but also
Balbus 2001 for the diabatic case with anisotropic conduction and
magnetic fields). It would be particularly interesting to contrast the
study of these generalized solutions with the results for averaged
flow properties emerging from simple hydrodynamical and MHD numerical
simulations (e.g. McKinney \& Gammie 2002; Proga 2005), in an effort
to make the best use of these two complementarity approaches to the
black hole accretion problem.

If conduction is generally important in hot accretion flows around
black holes, it could have additional interesting consequences besides
the ones illustrated in our specific solutions. For instance, if the
hot flow residing inside the capture radius around a black hole is
able to heat up the ambient gas outside of that capture radius, the
Bondi rate of gas capture ($\propto c_s^{-3}$) may be effectively
reduced. While Gruzinov (1998) invoked ``turbulent'' heat flows to
achieve this, the weakly-collisional conditions discussed in \S2 for
several galactic nuclei suggest that this could be done by
``microscopic'' conduction itself. More detailed models are therefore
needed to address various interesting issues related to conduction, to
understand better how conduction may affect the structure and
properties of hot accretion flows and to determine how these effects
may relate to the observational characteristics of dim accreting black
holes.

\acknowledgements
We are grateful to S. Balbus, A. Beloborodov, W. Dorland,
J.-P. Lasota, R. Narayan, E. Quataert and E. Spiegel for useful
discussions and comments. We also thank M. A. Morse for assistance in
producing the mass flux color figure. This research was supported in
part by the National Science Foundation under Grant No. PHY99-07949
(at KITP).

\appendix
  \begin{center}
    {\bf APPENDICES}
  \end{center}

\section{Prescriptions for Conduction}

In \S3, we have seen that a saturated form of conduction scales in
such a way that it permits 1D radially self-similar
solutions. Consequently, we first attempted to find 2D radially
self-similar solutions with a saturated conduction flux written as
\begin{equation}
{\bf F_{{\rm s}}}=\, -5\Phi_{{\rm s}} \rho c_s^3
\frac{{\bf \nabla} T}{|{\bf \nabla} T|},
\end{equation}
which depends only on the direction of the local temperature gradient,
not its magnitude.  From taking minus the divergence, we obtained the
following conduction term in the energy equation:
\begin{eqnarray}
Q_{{\rm s}}(\theta) &=& 5\Phi_{{\rm s}}
\Big\{\frac{P(\theta)c_{s}(\theta)}{\sqrt{1+\left(\frac{d}{d\theta}
\log c_{s}^{2}(\theta)\right)^{2}}}
\nonumber\\
\qquad\qquad\qquad\qquad
& &+\frac{1}{\sin\theta}\frac{d}{d\theta}\left[
\sin\theta\frac{P(\theta)c_{s}(\theta)}{\sqrt{1+\left(\frac{d}{d\theta}
\log c_{s}^{2}(\theta)\right)^{2}}}\frac{d}{d\theta}
\log c_{s}^{2}(\theta)\right]
\Big\},
\end{eqnarray}
within the framework of our self-similar solutions.

We have found this prescription to be numerically unstable, or at
least difficult to work with. This may be due to the explicit coupling
between the radial and polar conduction terms that appears within this
prescription (which does not seem well suited to our separable,
radially self-similar solutions). It may also be caused by the
structure of the mathematical operator resulting from this
prescription: it no longer acts as a diffusion process.

Consequently, in our 2D treatment (in \S4), we returned to a standard
diffusive operator for conduction, with a conductivity coefficient
scaled with radius to preserve self-similarity. This results in the
following conduction term in the energy equation (see also
Eq.~[\ref{energy}]):
\begin{equation}
Q_{{\rm d}} (\theta)=\, \lambda_{0}
\left[c_{s}^{2}(\theta)+\frac{1}{\sin\theta}\frac{d}{d\theta}
\left(\sin\theta \frac{d c_{s}^{2}(\theta)}{d\theta}\right)\right].
\end{equation}

From a comparison of these two prescriptions, it is clear that, in the
limit of strong conduction, when latitudinal profiles of most flow
quantities become nearly flat, the two conduction parameters,
$\Phi_{\rm s}$ and $\lambda_{0}$, are simply related to each by
$\lambda_{0} \approx \,5\rho(\theta)c_{s}(\theta)\,\Phi_{{\rm s}}$.
Using the simple scalings of Appendix B, the relation becomes
\begin{equation}
\lambda_{0}\approx\,\frac{5\sqrt{10}}{6\pi}\frac{\dot
m}{\alpha}\,\Phi_{{\rm s}} = 8.4 \,\Phi_{{\rm s}},
\end{equation}
where we have used $\dot m=1$, $\alpha=0.1$ for the last equality.
This correspondence is helpful in showing that, in all our 2D
solutions, the non-rotating limit is reached for $\Phi_{{\rm s}}$
equivalent values which are $\ll 1$, in good agreement with our 1D
results (in \S3). In other words, even in 2D solutions, a level of
conduction well below that of the (non-relativistic) saturated
conduction flux is sufficient to alter the structure and properties of
hot accretion flows substantially.

\section{Maximum Value of the Conduction Coefficient}

As we have seen in \S4, for a given value of the flow parameter
$\epsilon^{\prime}$, there is a maximum value of the conduction
parameter, $\lambda_{0}^{\prime}$, at which the solutions reach the
non-rotating, Bondi-like limit.  No solutions of physical significance
exist beyond this maximum value since they would require imaginary
values of $\Omega$, as we shall now establish.
Let us evaluate Eqs.~(\ref{moma}--\ref{energy}), at the flow equator,
using adequate boundary conditions, in the limit of low viscosity
($\alpha v \ll 1$), small radial velocity ($v^2 \ll 1$) and under the
assumption that second order derivatives with respect to $\theta$
become small (which is generally accurate for solutions approaching
the non-rotating limit). We obtain:
\begin{eqnarray}
\frac{2}{5} (1-\Omega^{2}),
\label{momae} &\approx &c_{s}^{2}\\
 (\frac{v_r}{c_{s}^{2}}+\frac{3}{2}\alpha)\Omega &\approx&0,
\label{momce}\\
-\frac{3}{2}\epsilon ^{\prime} v_r -\frac{\lambda_{0} ^{\prime}}{\rho}
 & \approx &\frac{9}{4}\alpha\Omega^{2}.
\label{energye}
\end{eqnarray}
The first equation establishes that, in the strong conduction limit,
$c_{s}^{2}\approx \frac{2}{5}$.  The second equation yields, in
combination with the first, $v_r \sim -\frac{3}{5} \alpha$, in the
same limit.  To be exact, this is not the limit $\Omega \approx 0$,
but rather $d^{2}\Omega/d\theta^{2}\approx 0$. The actual non-rotating
limit occurs at a slightly lower value of $v_r \lta -\frac{3}{5} \alpha$.
The third equation places the explicit constraint on $\lambda_{0}
^{\prime}$ for the non-rotating limit, since the left-hand side must
be positive for the value of $\Omega$ to be real.  This fixes the
maximum value of the conduction parameter $\lambda_{0}$ to
\begin{equation}
\lambda_{0,{}\max} (\dot m, \epsilon) \approx \frac{3}{4\pi}\dot
m\epsilon =0.24 \dot m\epsilon,
\label{Lmax}
\end{equation}
where we have used the crude approximation that momentum ($\rho v_r$)
profiles are uniform along $\theta$, which results in
$\rho(\theta)\sim 5\dot m /(6\pi \alpha)$.  This estimate of the
maximum value of $\lambda_{0}$ agrees well with our numerical results
(e.g. in Fig.~\ref{fig4}).

\section{Comparison with Anderson (1987)}

Our results on 2D hot accretion flows with conduction share strong
similarities with those of Anderson (1987). First, let us mention that
we have succeeded in reproducing many of the results reported by this
author, using the same prescription for conduction ( i.e. with a
conduction coefficient which scales not only with radius, but also
with the local flow viscosity and density).  With this prescription,
the conduction term in the energy equation becomes:
\begin{equation}
Q_{{\rm A}} =\, \alpha \lambda_{1} \left[
P(\theta)c_{s}^{2}(\theta)+ \frac{1}{\sin\theta}
\frac{d}{d\theta}\left(\sin\theta P(\theta)\frac{d
c_{s}^{2}(\theta)}{d\theta}\right)\right].
\end{equation}

Returning to our analysis, there are still differences between our
work and that of Anderson (1987), both in the formalism and in the
results. Anderson works in the limit of ``weak shear viscosity,'' with
several viscous terms dropped from the dynamical equations (as
compared to ours). He adopts a polytropic equation of state, a
prescription for viscosity which differs from Shakura-Sunyaev and
there is no inertial term in his radial momentum equation.  As a
result, his set of differential equations is reduced to a third order
system, as opposed to a seventh order one in our case. Despite these
technical differences, our results are qualitatively consistent with
many of his conclusions. Differences appear mostly at the quantitative
level, for instance in the rates of mass outflow (which appear a few
times larger in our solutions). One qualitative feature which has
apparently been missed by Anderson (1987) is the existence of
solutions with conical outflow geometries in sub-regions of the
parameter space.

From this comparison, we can safely conclude that many of the
qualitative features exhibited by our solutions should be relatively
robust to changes in the various prescriptions adopted (e.g.
conduction and viscosity).

\clearpage

\begin{table*} 
\begin{tabular}{lccccc}
\hline
\\
Nucleus & $n_{1''}$ & $T_{1''}$ & {$R_{1}$} & {$l_1/R_1$} & {$l_1/R_{\rm cap}$} \\
 & (cm$^{-3}$) & ($10^7$~K) & (cm)& & \\
\\
\hline
\\
Sgr A*    & $100$ &  $2.3$  & $1.3 \times 10^{17}$ & $0.4$ & $0.4$\\
NGC 1399  & $0.3$ &  $0.9$  & $3.1 \times 10^{20}$ & $0.009$ & $0.02$\\
NGC 4472  & $0.2$ &  $0.9$  & $2.5 \times 10^{20}$ & $0.016$ & $0.07$\\
NGC 4636  & $0.07$&  $0.7$  & $2.2 \times 10^{20}$ & $0.032$ & $0.6$\\
M87       & $0.17$&  $0.9$  & $2.7 \times 10^{20}$ & $0.018$ & $0.02$\\
M32       & $0.07$&  $0.4$  & $1.2 \times 10^{19}$ & $0.2$   & $1.3$
\\
\hline
\end{tabular}
\hspace{8.5cm} \caption{Observational Constraints on Collisionality in
Nearby Galactic Nuclei \label{tab:one}}
\end{table*}

\clearpage 


\end{document}